\documentclass[prl,preprintnumbers,twocolumn,aps,superscriptaddress]{revtex4}
\usepackage{graphicx,}
\usepackage{graphicx,dcolumn,bm,epsfig,amsmath,amssymb,textcomp,}
\usepackage{float}

\begin{document}

\title{Effects of oscillating spacetime metric background on a complex scalar field and formation of 
topological vortices}

\author{Shreyansh S. Dave}
\email{shreyanshsd@imsc.res.in}
\affiliation{The Institute of Mathematical Sciences, Chennai 600113, India}
\author{Sanatan Digal}
\email{digal@imsc.res.in}
\affiliation{The Institute of Mathematical Sciences, Chennai 600113, India}
\affiliation{Homi Bhabha National Institute, Training School Complex,
Anushakti Nagar, Mumbai 400085, India}

\begin{abstract}
We study the time evolution of a complex scalar field in the symmetry broken phase in the
presence of oscillating spacetime metric background. In our (2+1)-dimensional simulations,
we show that the spacetime oscillations can excite an initial field configuration, which 
ultimately leads to the formation of topological vortices in the system. At late times, 
field configuration achieves a {\it disordered state}. A detailed study of the momentum 
and frequency modes of the field reveals that these field excitations are driven by the 
phenomenon of {\it parametric resonance}. In extremely high frequency regime where frequency 
of spacetime oscillations is much larger than the field-mass, the formed vortices are not 
topological in nature. Interestingly in this regime, for a suitable choice of parameters of 
the simulation, we observe a persistent lattice structure of vortex-antivortex pairs. We 
discuss applications of our study to the dynamics of interior superfluidity of neutron stars 
during binary neutron star mergers, in generation of excitation in ultralight axion-like 
field near a strong gravitational wave source, etc.
\end{abstract}

\maketitle

\noindent

\section{I. Introduction}

Topological defects exist in systems ranging from condensed matter to the early Universe 
\cite{zrk1,rjnt}. They exist in systems which have topologically non-trivial order 
parameter space (or vacuum manifold) \cite{mermin}. There are many condensed matter systems, 
e.g. superfluid, superconductor, nematic liquid crystal, etc., which have topologically 
non-trivial order parameter space. For example, the superfluid phase of $^4$He has circle 
$S^1$ as an order parameter space, for which the {\it fundamental group} is non-trivial, 
i.e. $\pi_1(S^1)$=$\mathcal{Z}$ \cite{mermin}. Therefore in this system, topological vortices 
could exist. Topological vortices are characterized by different elements of the fundamental 
group.

There are various ways by which topological defects can be formed in a physical system. These 
defects can form during spontaneous symmetry breaking phase transition, formation of which is 
described by the Kibble-Zurek mechanism \cite{zrk1,kbl}. However, in the presence of external 
influence, this mechanism needs some modifications \cite{skyrm,bias}. There are other methods also 
by which these defects can be formed. For example, in superfluid $^4$He, the rotation of vessel 
within a range of angular velocity leads to the formation of vortex lattice \cite{landau,tilley}. 
Similarly, in type-II superconductor, an external magnetic field within a range of field 
strength leads to the formation of flux-tube lattice \cite{landau,tilley}.

In refs.\cite{resodef,kink}, an interesting possibility of production of topological defects 
is investigated in $\Phi^4$ theories. In these studies, defect-antidefect pairs have 
been produced under oscillating temperature of the effective potential with frequency close 
to mass of the field. This leads to the {\it parametric resonance} of the field, under which 
the field achieves excitations leading to the formation of vortex-antivortex pairs in 
U(1)-theory \cite{resodef} and kink-antikink pairs in $\mathcal{Z}_2$-theory \cite{kink}.

In Bose-Einstein condensate (BEC) of ultracold atoms, a similar method of formation of 
superfluid vortices has been studied \cite{oscBEC1,oscBEC2,resoBEC1}. In these studies, 
vortex-antivortex pairs have been produced by periodically varying trapping potential of 
condensate. These studies suggest that under these periodic perturbations, time dependent 
excitations arise in the condensate modifying the coherent wavefunction of BEC. These 
excitations lead to the formation of vortex-antivortex pairs in the system. Subsequently, 
a tangled network of vortices is formed in the system, indicating a transition to a state 
of quantum turbulence. Ultimately, the system reaches a disordered state where BEC is 
completely destroyed. In these studies, the oscillation frequency of trapping potential 
is considered to be 200 Hz, and the time scale for the whole process is few tens of 
milliseconds. This time scale depends upon the amplitude and frequency of oscillations of 
trapping potential (i.e. rate of injected energy) and the energy of vortex configuration 
in the system. 
 
Following above studies in refs.\cite{resodef,kink,oscBEC1,oscBEC2,resoBEC1}, it is an 
obvious question to ask if spacetime oscillations can also induce parametric resonance or 
large oscillations in a field. Such a question is relevant for many systems, for example, 
ultralight axion-like field coupled to gravitational waves, neutron star superfluidity 
under time dependent tidal deformations, etc. During a binary neutron stars (BNS) merger, 
the orbiting neutron stars exert a time dependent tidal force on each other \cite{bns1}. 
The frequency and amplitude of this tidal force keep on increasing with time, and become 
maximum towards the end of merger. Under such tidal force, depending upon the {\it tidal 
deformability}, neutron star undergoes time dependent deformations. These tidal 
deformations can couple to the condensate of  interior superfluid of neutron star and may 
lead to the turbulence (formation of tangled vortices) as discussed in 
refs.\cite{oscBEC1,oscBEC2} (for the discussion on the possibility of superfluidity inside 
neutron star, see refs.\cite{ns1,ns2,krishna}). A similar kind of scenario, i.e. occurrence 
of superfluid turbulence, superfluid to normal fluid transition, and glitch (or anti-glitch) 
in neutron stars during BNS merger has been discussed in ref.\cite{proposal} as well. 

It has been studied in ref.\cite{recentMerger} that due to the tidal force during the merger, 
energy pumped into the system is so high that the (local) temperature and density of neutron 
stars can reach up to 100 MeV and few times of nuclear saturation density, respectively. 
Therefore, in such a situation neutron superfluidity can be destroyed (in some local regions) 
even before the completion of merger as the transition temperature for such superfluidity is 
somewhere in between 0.1 to 5 MeV \cite{nsSFtc}. Therefore the generation of excitations, 
vortex-antivortex pairs, and quantum turbulence in superfluid during the merger are expected 
to occur in the intermediate stage of evolution of neutron stars. It is also possible that the 
condensate excitation increases so much that the neutron star superfluidity is destroyed even 
before the transition temperature is reached (as discussed in studies \cite{oscBEC1,oscBEC2}). 
Certainly, the superfluid phase of quark matter (e.g. color-flavor locked phase) in the inner 
core of neutron stars can survive for a longer time as it has transition temperature $\sim$100 
MeV \cite{krishna}.

To study the time evolution of interior superfluidity of neutron star under such tidal 
deformations, one must solve the relativistic Gross-Pitaevskii equation, known as the 
Gross-Pitaevskii-Anandan equation \cite{anandan}, in the background of spacetime metric of star 
with a time dependent perturbation appropriate for the BNS merger system. This also can be done 
by solving non-linear Klein-Gordon equation in the presence of an appropriate metric background 
as superfluidity can also be described in the field theoretical framework, where BEC is generally 
characterized by spontaneous breaking of U(1) global symmetry for a complex scalar field 
\cite{introSF,relSF1}.

In this paper, we study the effects of spacetime oscillations on the evolution of a complex scalar 
field. This may shed some light on the time dynamics of interior superfluidity of neutron star during 
merger. However, in this work we are not doing phenomenology for any particular system. Rather, the 
present work is completely formal, where we show for the first time that a complex scalar field can be 
excited and topological vortices can be formed under the spacetime oscillations. Therefore, here instead 
of considering a spacetime metric of star, for simplicity, we take Minkowski metric with a periodic 
perturbation in time. This simplification helps to give a clear understanding of the effect of metric 
oscillations on the field and production of topological vortices. In this metric background, we 
numerically solve the non-linear Klein-Gordon equation in (2+1)-dimensions for a complex scalar field 
with an initial field configuration given by, $\Phi (x,y)$=$\Phi_0$+$\delta \Phi (x,y)$, where $\Phi_0$ 
is the vacuum expectation value (VEV) of U(1) symmetry broken effective potential and $\delta \Phi (x,y)$ 
represents small 
fluctuations of field about the VEV. Here, the initial field fluctuations have been considered as the 
spacetime oscillations couple with the field equation only through the spacetime derivatives of the 
field; see equation of motion below.

We see that under the spacetime oscillations, the field undergoes large amplitude oscillations for a wide range of 
frequencies of spacetime oscillations, which eventually leads to the formation of vortex-antivortex pairs. In 
the low frequency regime (frequency smaller than the mass of field) mainly transverse modes (Goldstone modes) 
arise, while in the high frequency regime, longitudinal modes also get generated dominantly. A detailed analysis 
shows that there is a correspondence between the frequency of spacetime oscillations and momentum of time growing 
field-modes. This suggests that the field undergoes {\it parametric resonance} under the spacetime oscillations. 

We mention that our present study is fundamentally different from the previous studies in refs.\cite{resodef,kink}, 
even though the underlying physics for the generation of topological defects is same, i.e. {\it parametric resonance}.
In previous studies oscillating temperature couples to the magnitude of the field. Therefore, in order to observe 
parametric resonance, the frequency of temperature oscillations must be close to the mass of longitudinal modes of 
the field. Whereas in present case, the spacetime oscillations couple to the gradient of field. This makes a direct 
coupling between the frequency of spacetime oscillations and the momentum of field-modes present in the initial field
configuration. Therefore by satisfying the resonance condition, some specific field-modes grow exponentially with 
time. This can occur, as long as the appropriate momentum-modes of the field are present in the system. Thus, in 
present case mass of the field does not set the lowest frequency cut-off to induce parametric resonance, hence
vortex-antivortex pairs could be formed for a wide range of frequencies of spacetime oscillations. However, for this 
phenomenon, the lowest frequency cut-off is set by the system size due to finite size effects. Our results suggest 
that the frequency of spacetime oscillations must be greater than 2/(system size) to induce the parametric resonance
of the field.

In the case of neutron stars during BNS merger, the accessible frequencies of spacetime oscillations are 
many orders of magnitude smaller than the mass of condensate field of interior superfluidity ($\sim$ 
0.1 - 5.0 MeV). Therefore in this case, even the transverse-modes of condensate field may get difficulties 
to grow due to finite size effects. Our present results suggest that in a typical neutron star of radius 
$\sim$10 km, to generate condensate excitations, the frequency of spacetime oscillations must be greater than 
$\sim$30 kHz. This required frequency is way beyond the reach of any known BNS merger systems, which could 
generate maximum frequency of spacetime oscillations up to $\sim$1 kHz. Therefore, it is difficult to generate 
condensate excitations of neutron star superfluidity, under the phenomenon of parametric resonance, during BNS 
merger. 

However, this has to be investigated in detail that whether there is any other method by which 
condensate excitation in neutron star superfluidity can be generated during mergers. We mention 
that in the case of BEC of ultracold atoms, a 200 Hz frequency of oscillations of trapping potential 
is sufficient to excite the condensate with system size of only few hundred micrometers in a few tens 
of milliseconds time \cite{oscBEC1}. Therefore, it is not very unrealistic to expect the generation 
of such condensate excitations and formation of vortex-antivortex pairs in the interior 
superfluidity of neutron star during BNS merger. This requires a detailed investigation, which we 
will try to pursue in the future.

This paper is organized as follows. In section-II, we derive the equation of motion for a complex scalar field 
in the presence of oscillating spacetime metric background. In section-III, we outline the details of our 
numerical simulations. Then we present our simulation results in section-IV. Finally, we briefly summerize our 
work and discuss future directions in section-V.

\section{II. Equation of Motion}

To study the effects of spacetime oscillations on a complex scalar field, we consider the spacetime 
metric as a periodic perturbation on the top of Minkowski metric. We consider the {\it inverse} spacetime 
metric as, $g^{\mu \nu}\equiv diag \Big(-1,1-\varepsilon \sin \big(\omega (t-z)\big),1+\varepsilon \sin \big(\omega (t-z)\big),1\Big),$
where $\varepsilon$ ($<1$) and $\omega$ are the amplitude and frequency of spacetime oscillations, respectively; 
$(t,x,y,z)$ are spacetime coordinates. Note that, this form of the metric has been chosen just for a simplicity in 
the equation of motion of the field, though any periodic time-dependent metric can be taken for our study. The 
action of the complex scalar field on the spacetime manifold with the given metric is \cite{carroll,relSF1} 
\begin{equation}
 S=\int d^4x \sqrt{-g}\Big[-\frac{1}{2}g^{\mu \nu} \partial_{\mu}\Phi^* \partial_{\nu} \Phi - V(\Phi^* \Phi)\Big],
 \label{eq.Lagrangian}
\end{equation}
where $g$=$det(g_{\mu \nu})$=$-\Big($1$-$$\varepsilon^2\sin^2\big(\omega (t$$-$$z)\big)\Big)^{-1}$, $\Phi$=$\phi_1$+ $i\phi_2$, 
$\Phi^*$=$\phi_1$$-$$i\phi_2$; $\phi_1$ and $\phi_2$ are the real scalar fields. We consider the symmetry 
breaking effective potential
\begin{equation}
 V(\Phi^* \Phi)= \frac{\lambda}{4} \Big(\Phi^* \Phi - \Phi_0^2 \Big)^2,
 \label{potential}
\end{equation}
where $\lambda$ is the self-interaction coupling parameter of the field, and $\Phi_0$ is the VEV of the effective 
potential. With this, the mass of longitudinal component of the field is given by $m_\Phi$=$\Phi_0 \sqrt{2\lambda}$. 
The equations of motion for ($\phi_1,\phi_2$) fields are \cite{carroll,relSF1}
\begin{equation}
 \Box \phi_i-\frac{dV}{d\phi_i}=0, 
\end{equation}
where $i=1,2$. The covariant d'Alembertian is given by
\begin{equation}
 \Box \phi_i = \frac{1}{\sqrt{-g}}\partial_{\mu}\Big(\sqrt{-g} g^{\mu \nu} \partial_{\nu}\phi_i \Big).
\end{equation}
Therefore in the expanded form, the field equations become
\begin{equation}
 \begin{split}
  -\frac{1}{2} \frac{\varepsilon^2 \omega \sin \big(2\omega (t-z)\big)}{\Big(1-\varepsilon^2 \sin^2 \big(\omega (t-z)\big)\Big)} 
  \Bigg(\frac{\partial \phi_i}{\partial t}
  +\frac{\partial \phi_i}{\partial z}\Bigg)
 -\frac{\partial^2\phi_i}{\partial t^2}\\ + 
 \Big(1-\varepsilon \sin \big(\omega (t-z)\big) \Big)\frac{\partial^2\phi_i}{\partial x^2}
 +\Big(1+\varepsilon \sin \big(\omega (t-z)\big) \Big)\frac{\partial^2\phi_i}{\partial y^2}\\ + \frac{\partial^2\phi_i}{\partial z^2}
 - \lambda \phi_i \Big(\phi_1^2+\phi_2^2-\Phi_0^2 \Big)=0.
\end{split}
\end{equation}
For the simplicity of solving it numerically, (i) we assume that there is no variation of the field $\Phi$ along 
$z$-direction, and (ii) we look at the solution of the field only in the $z$=0 plane. With these simplifications, 
the above equations reduce to
\begin{equation}
 \begin{split}
  -\frac{1}{2} \frac{\varepsilon^2 \omega \sin (2\omega t)}{\big(1-\varepsilon^2 \sin^2 (\omega t)\big)} 
  \frac{\partial \phi_i}{\partial t}  -\frac{\partial^2\phi_i}{\partial t^2} + 
 \big(1-\varepsilon \sin (\omega t) \big)\frac{\partial^2\phi_i}{\partial x^2}\\
 +\big(1+\varepsilon \sin(\omega t) \big)\frac{\partial^2\phi_i}{\partial y^2} 
  - \lambda \phi_i \Big(\phi_1^2+\phi_2^2-\Phi_0^2 \Big)=0.
\end{split}
\label{eom}
\end{equation}
It is very clear from the above equations that the spacetime oscillations only couple to the derivatives of the 
field, where the coefficient of the first order time derivative term oscillates with a mixture of frequencies 
$\omega$ and $2\omega$. Under the spacetime oscillations, gradient of the field oscillates, which must induce 
oscillations in the field via the {\it parametric resonance} with different possible oscillation frequencies. 
These field oscillations must be modulated by the steep rise of the effective potential in the longitudinal 
direction. The parametric resonance of the field eventually leads to the formation of vortex-antivortex pairs in 
the system. Our simulation results validate these expectations, which we present in the following sections.

\section{III. Simulation Details}

To perform the simulation, we discretize the $xy$-plane into 200$\times$200 lattice points with the lattice spacing of 
$\Delta x$=$\Delta y$=0.01 $\Lambda$, which gives the total lattice size $L$=$2.0~\Lambda$ in each direction (here, 
$\Lambda$ is a unit of length and time, which we have not specified, as in this work we are not presenting phenomenological 
prediction for any system). $\Delta x$ and $\Delta y$ should be smaller than inverse of the field-mass, i.e. 
$\Delta x,\Delta y < m_\Phi^{-1}$, so that a reasonable spatial variation of the field can be resolved. The time is 
discretized in the time step of $\Delta t$=0.005 $\Lambda$. Note that in the equation of motion (Eq.(\ref{eom})), the
discretization in the time-axis works only if the condition $\Delta t<\omega^{-1}$ is satisfied. In our simulation, we study 
the effects of spacetime oscillations, in particular, the role of parameters $\varepsilon$ and $\omega$ on the field 
evolution. To describe the effects, for most of the simulations, we consider $\varepsilon$=0.4 and $\omega$=100 
$\Lambda^{-1}$. We take the parameters of the effective potential as $\Phi_0$=10 $\Lambda^{-1}$ and $\lambda$=40. We also 
study the effects of changing these parameters on simulation results.

In Eq.(\ref{eom}), the field evolution is coupled with spacetime oscillations through second order spatial derivative and first 
order time derivative of the field. Therefore initial field configuration can respond to the spacetime oscillations only when 
fluctuations of the field (spacetime gradient) are present in the system, i.e. iff the field configuration is not in the 
complete minimum energy configuration. These fluctuations can naturally arise due to the presence of thermal and/or quantum 
fluctuations in the system. For simplicity, we consider fluctuations only in the initial field configuration, i.e. we take 
$\Phi(x,y)$=$\Phi_0$+$\delta \Phi(x,y)$ at time $t$=0 and further evolve the field with Eq.(\ref{eom}).

In our simulation, for the initial field configuration, we have considered small spatial variations in $\phi_1$ and $\phi_2$ 
fields such that the magnitude of field $\Phi$ remains $\Phi_0$, i.e the field is taken at the minima of the effective potential 
(Eq.\ref{potential}) and only transverse fluctuations of the field have been considered for the simulation. However, we have 
seen that considering fluctuations in the magnitude of the field does not change any qualitative aspect of the simulation 
results. More specifically, we have taken $\phi_1(x,y,t$=$0)$=$\Phi_0\sqrt{1-\alpha_{_{xy}}^{2}}$ and 
$\phi_2(x,y,t$=$0)$=$\Phi_0 \alpha_{_{xy}}$, where $\alpha_{_{xy}}$ varies randomly from point to point on the lattice in the 
range [$-\beta,\beta$]; here 0$<$$\beta$$\ll$ 1. This initial field configuration has a non-zero spatial gradient of both  
fields. In our simulation, we have taken $\beta$=0.2. We have seen that choosing different values of $\beta$ does not change our
results qualitatively. We evolve this initial field configuration by solving Eq.(\ref{eom}) using the second-order Leapfrog method, 
and considering periodic boundary condition along the spatial directions. By changing lattice size and boundary conditions, we 
have checked that the reflection of field fluctuations from the boundary has a negligible effect on our simulation results.

\section{IV. Simulation Results}

\subsection{A. Parametric resonance of field under spacetime oscillations}

\begin{figure}
\includegraphics[width=1.\linewidth]{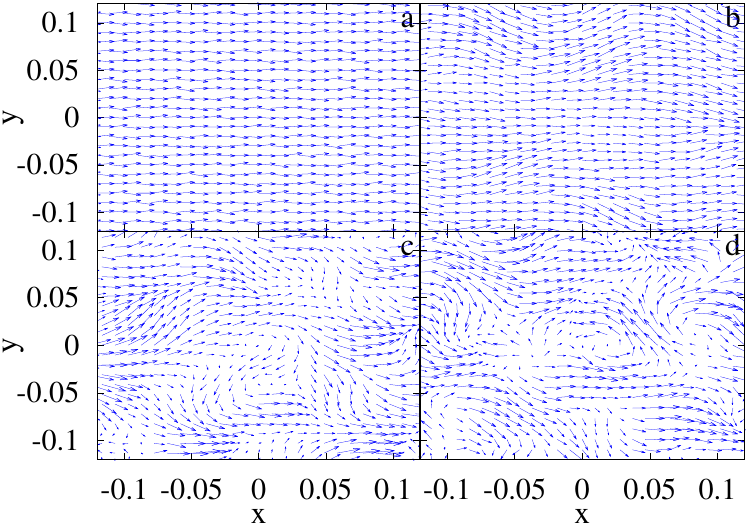}
 \caption{Figure shows the vector plots of field configuration in physical space ($xy$-plane) at four different times of field 
 evolution. The parameters of simulation are $\varepsilon$=0.4, $\omega$=100 $\Lambda^{-1}$, $\Phi_0$=10 $\Lambda^{-1}$, and
 $\lambda$=40.  Plots (a),(b),(c), and  (d) correspond to the field configurations at times $t$=0, 1.35, 1.7, and  
 $1.8~\Lambda$, respectively. In plot(d) vortex-antivortex pairs have been formed. Roughly at time  $t$=2.5 $\Lambda$, the field 
 configuration achieves a  {\it disordered state}.}  
 \label{fig1}
\end{figure}

We now present our simulation results. In Fig.\ref{fig1}, we show the vector plots of field configuration in the 
physical space ($xy$-plane) at four different times of the field evolution.  The $x$ and $y$ components of the vector field 
correspond to $\phi_1$ and $\phi_2$ fields, respectively. Note that the whole lattice area is 2.0$\times$2.0 $\Lambda^2$, 
while only a small portion of the lattice is shown in the figure. In this simulation, we have taken $\varepsilon$=0.4, 
$\omega$=100 $\Lambda^{-1}$, $\Phi_0$=10 $\Lambda^{-1}$, and $\lambda$=40. As mentioned earlier that, only transverse 
fluctuations of the field have been considered in the initial field configuration, for which phase of the field varies randomly 
from one lattice point to other within the range of $-0.2$ to $0.2~rad$ for the given value of $\beta$. Fig.\ref{fig1}(a) 
shows such an initial field configuration, in which the phase fluctuations of field can be clearly seen. We observe that 
roughly during the time interval of $0.5$ - $1.5~\Lambda$, the field itself starts performing oscillations in space and time 
with a significantly large amplitude; at time larger than this, the field dynamics becomes much more complicated. Fig.\ref{fig1}(b) 
shows the field configuration at time $t$=1.3 $\Lambda$. This figure clearly shows that there is generation of a systematic 
periodic variation (waves) of the field in the physical space, whose amplitude keeps on increasing with time within the time 
interval mentioned above. We observe that the wavelength of these waves is $\omega$ dependent; decreasing $\omega$ leads to increase 
in wavelength of these waves (see Fig.\ref{fig4} and Fig.\ref{fig5} in this regard). Depending upon the choice of $\omega$, these 
waves may contain transverse as well as longitudinal excitation of the field. Because of the generation of these excitations, the field 
acquires a significant energy in some parts of the physical space to climb the central barrier of the effective potential and fall 
into the opposite side of the vacuum manifold. This leads to the formation of vortex-antivortex pairs in the system. This process 
can be seen in Fig.\ref{fig1}(c) and Fig.\ref{fig1}(d) which are plotted at times $t$=1.7 $\Lambda$ and $t$=1.8 $\Lambda$, respectively. 
In this process, the vortex-antivortex pairs keep on forming and some of them annihilating with time, while the overall number of 
these pairs keeps on increasing. Roughly at the time $t$=2.5 $\Lambda$, fluctuations become so strong that the field configuration 
no longer remains in the {\it ordered state} and achieves a {\it disordered state} (not shown in the figure).

\begin{figure}
\includegraphics[width=1.\linewidth]{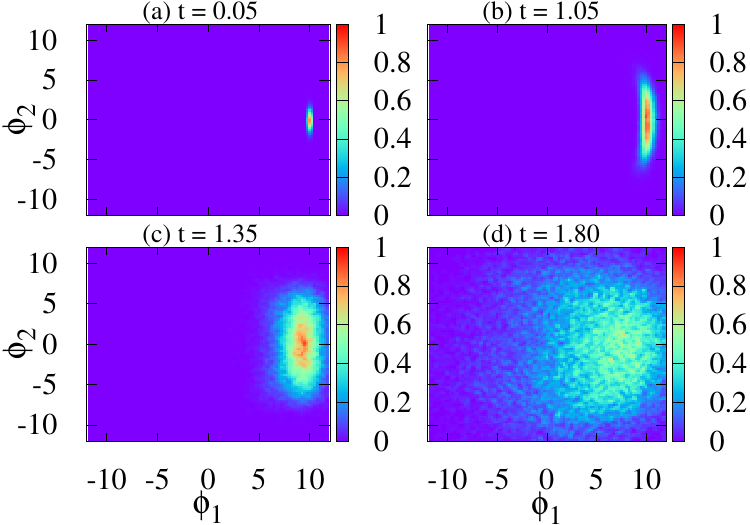}
 \caption{In the figure, the distribution of field has been plotted in field-space ($\phi_1 \phi_2$-plane) at four 
 different times of field evolution. The distribution has been normalized such that in each plot density varies 
 in between 0 and 1. Plots (a),(b),(c), and (d) correspond to the field distribution at times $t$=0.05, 1.05, 1.35, and 
 $1.8~\Lambda$, respectively. In plot(a), field distribution is highly localized about the chosen minima of the effective 
 potential, which becomes more spread in plot(b), where mostly transverse excitation has been generated. In plot(c) 
 longitudinal excitation also has been generated. At the time stage of plot(d), field flips to the opposite side of the 
 vacuum manifold leading to the formation of vortex-antivortex pairs.}
 \label{fig2}
\end{figure}

In Fig.\ref{fig2}, we plot distribution of the field in the $\phi_1 \phi_2$-plane (field-space) at four different 
times of the field evolution. In this figure, the field has been mapped from the physical space to the field-space, where 
the density of points has been indicated by colors. The distribution has been normalized such that in each plot density 
varies in between 0 and 1. The parameters of simulation are same as used in Fig.\ref{fig1}. Fig.\ref{fig2}(a) shows the 
field distribution at time $t$=0.05 $\Lambda$. This shows a highly localized distribution of the field around $\phi_1$$\approx$$\Phi_0$=10 
$\Lambda^{-1}$, $\phi_2$$\approx$0 due to our initial choice of field values. In this figure, the maximum height of 
the distribution is 2672. As field evolves, the distribution keeps on spreading around the initial distribution in the 
field-space. Fig.\ref{fig2}(b) shows the field distribution at time $t$=1.05 $\Lambda$ with maximum distribution height of 
316. One can clearly see that at this time, mostly the transverse excitation has been generated. Fig.\ref{fig2}(c) shows 
the stage of field distribution at time $t$=1.35 $\Lambda$ with distribution height of 88. At this time, both types of 
excitations, transverse as well as longitudinal, have been generated in the system. Note that Fig.\ref{fig2}(b) and 
Fig.\ref{fig2}(c) correspond to the time duration in which field oscillates with a significantly large amplitude, as shown 
in Fig.\ref{fig1}(b). These excitations ultimately lead to the formation of vortex-antivortex pairs in the system. The 
field distribution, at the time stage when these pairs have been formed, is shown in Fig.\ref{fig2}(d) at time $t$=1.8 
$\Lambda$ with maximum distribution height of 26. Subsequently, the field distribution spreads over the field-space 
symmetrically around $\phi_1$=$\phi_2$=0, the maximum extent of which becomes way beyond the VEV, signaling a {\it 
disordered state} of the field.   

\begin{figure}
  \includegraphics[width=1.0\linewidth]{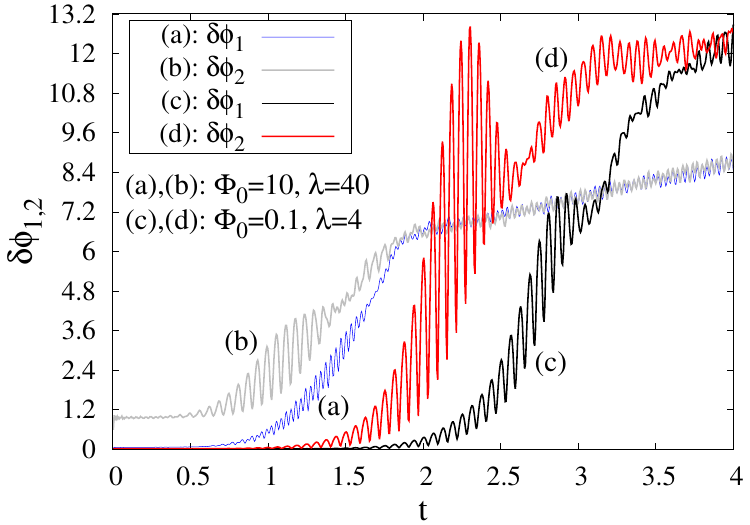}
  \caption{Figure shows the time evolution of fluctuations $\delta \phi_1$ and $\delta \phi_2$ of fields 
  $\phi_1$ and $\phi_2$, respectively, under the spacetime oscillations for two sets of parameters of the 
  effective potential. The parameters of spacetime metric are $\varepsilon$=0.4 and $\omega$=100 $\Lambda^{-1}$. 
  Plots (a)(blue) and (b)(gray) are the time evolution of $\delta \phi_1$ and $\delta \phi_2$ for $\Phi_0$=10 
  $\Lambda^{-1}$ and $\lambda$=40, while plots (c)(black) and (d)(red) are for $\Phi_0$=0.1 $\Lambda^{-1}$ and 
  $\lambda$=4, respectively. Figure clearly shows that in each case, the initial growth of $\delta \phi_2$ is 
  larger in comparison with $\delta \phi_1$; $\delta \phi_1$ and $\delta \phi_2$ ultimately become degenerate. 
  There are oscillations in plots (a) and (b) during the time interval of 0.5 - 1.5 $\Lambda$, which correspond 
  to the large amplitude field oscillations discussed in Fig.\ref{fig1}.} 
  \label{fig3}
\end{figure}

\begin{figure*}
  \includegraphics[width=0.268\linewidth]{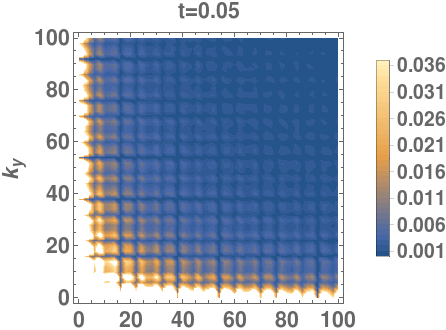}
   \includegraphics[width=0.238\linewidth]{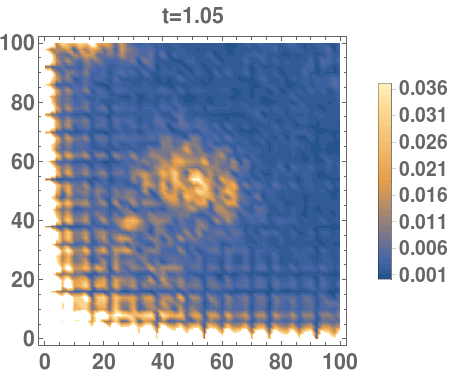}
     \includegraphics[width=0.238\linewidth]{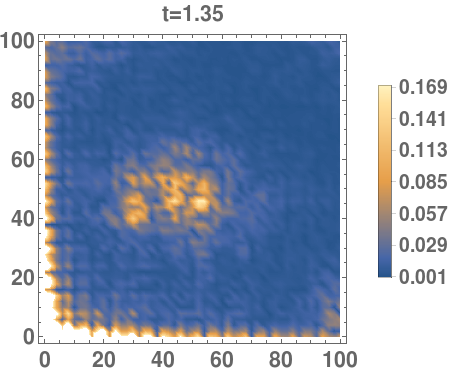}
       \includegraphics[width=0.238\linewidth]{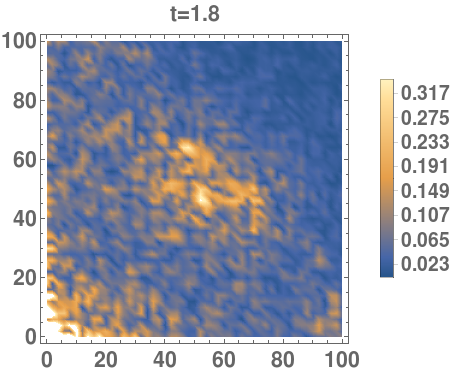}
         \includegraphics[width=0.272\linewidth]{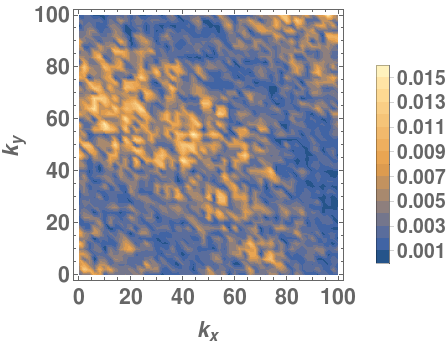}
             \includegraphics[width=0.236\linewidth]{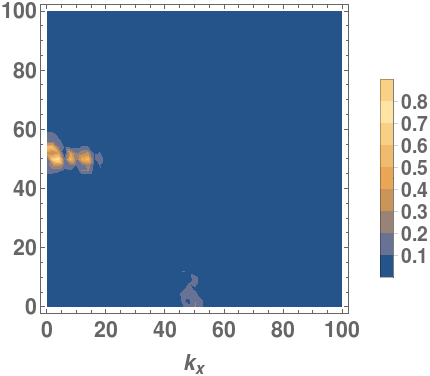}
               \includegraphics[width=0.236\linewidth]{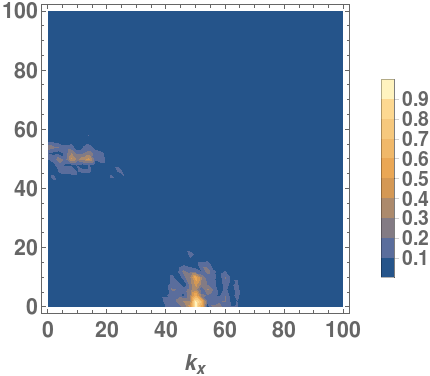}
                 \includegraphics[width=0.236\linewidth]{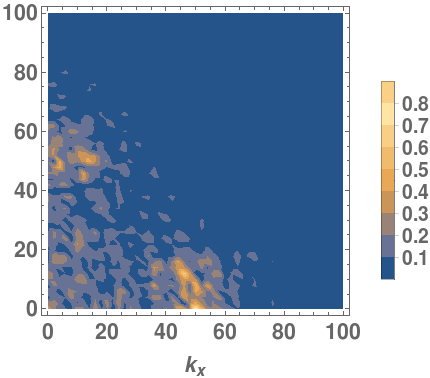}
  \caption{Figure shows the profile of modulus of field-modes $\lvert \tilde{\phi}_1(\vec{k},t)\lvert$ (upper panel) and 
  $\lvert \tilde{\phi}_2(\vec{k},t)\lvert$ (lower panel) at different times for $\omega$=100 $\Lambda^{-1}$, $\varepsilon$=0.4, 
  $\Phi_0$=10 $\Lambda^{-1}$, $\lambda$=40. From left to right, modes are plotted at time $t$=0.05, 1.05, 1.35, 1.8 $\Lambda$. 
  There is noticeable growth of $\lvert \tilde{\phi}_1(\vec{k},t)\lvert$ at $k_x$=15 $\Lambda^{-1}$, $k_y$=$\omega$ and 
  $k_x$$\approx$$k_y$$\approx$$\sqrt{\big(\omega^2-(\frac{m_\Phi}{\sqrt{2}})^2\big)/2}$ =54.77 $\Lambda^{-1}$, and growth of
  $\lvert \tilde{\phi}_2(\vec{k},t)\lvert$ at $k_x$=0, $k_y$=$\omega/2$ and $k_x$=$\omega/2$, $k_y$=0. The subsequent evolution 
  also leads to the growth of various other momentum-modes of both fields.} 
  \label{fig4}
\end{figure*}

\begin{figure*}
  \includegraphics[width=0.245\linewidth]{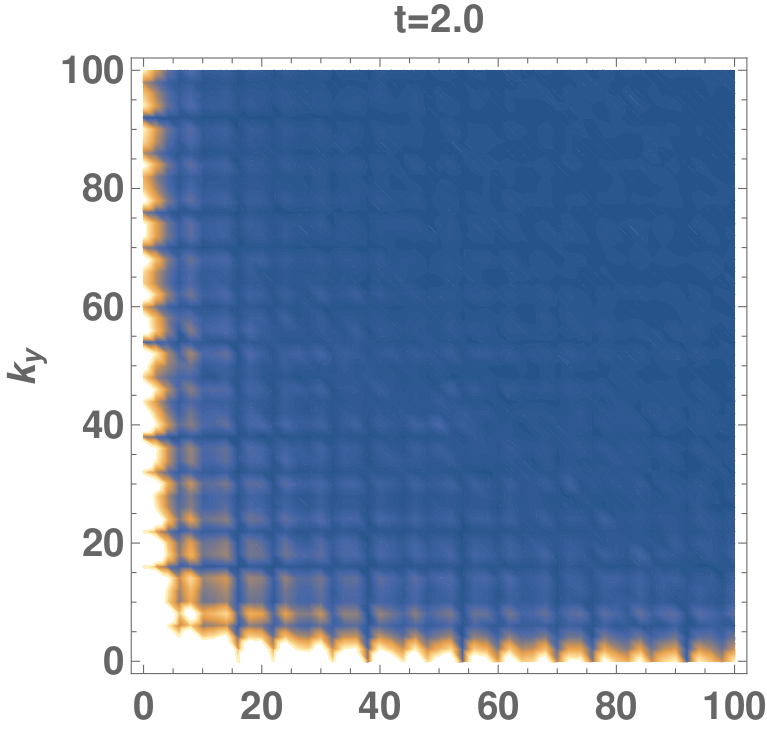}
   \includegraphics[width=0.225\linewidth]{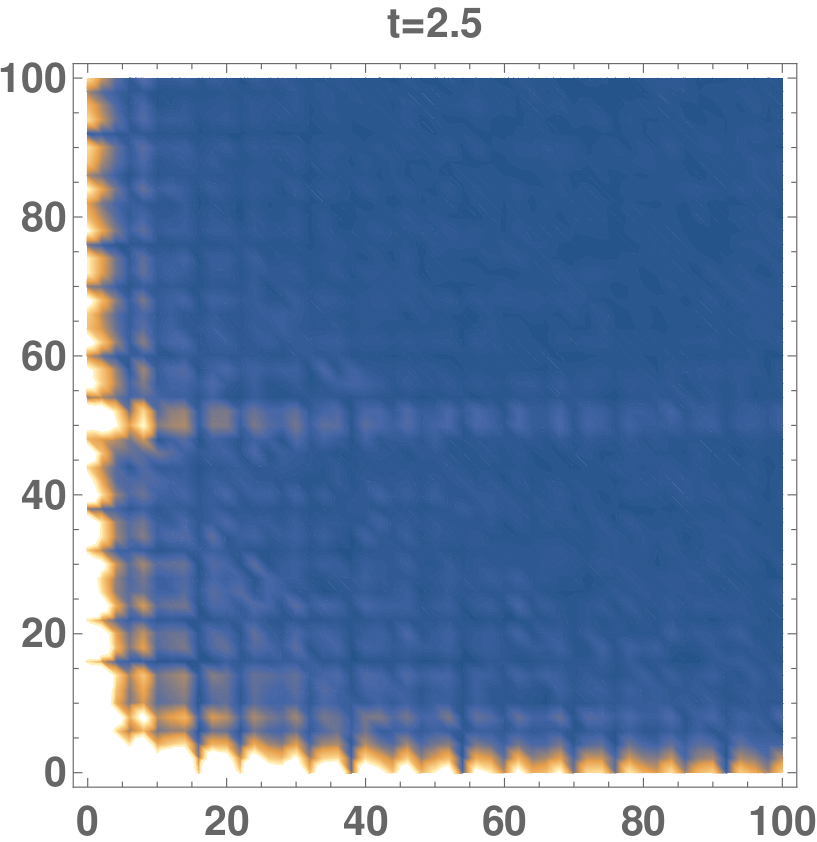}
    \includegraphics[width=0.225\linewidth]{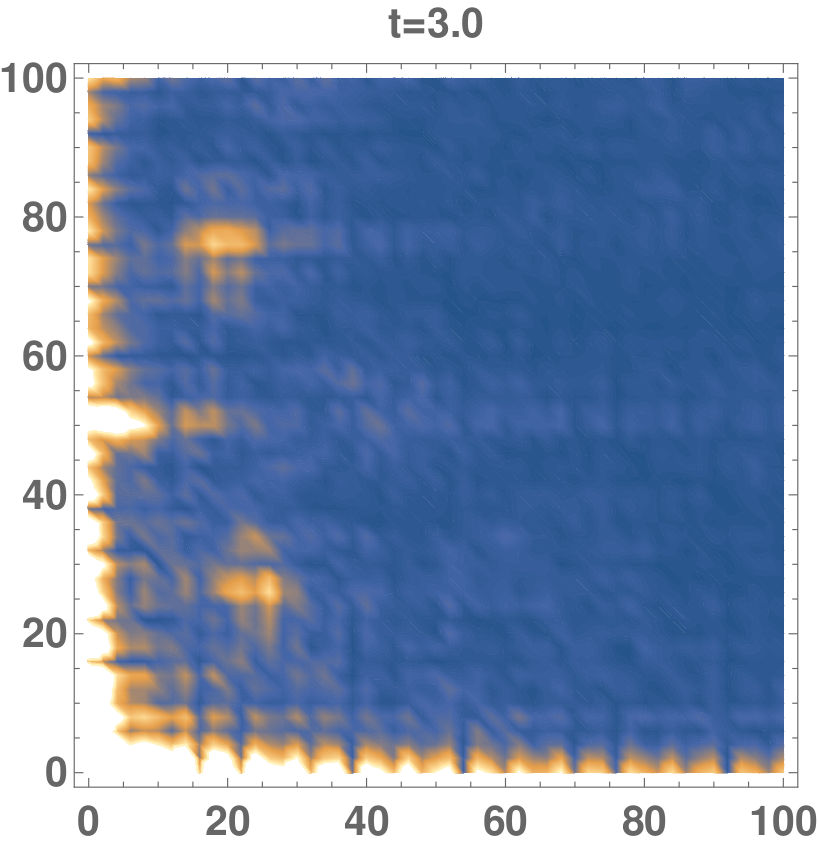}
     \includegraphics[width=0.225\linewidth]{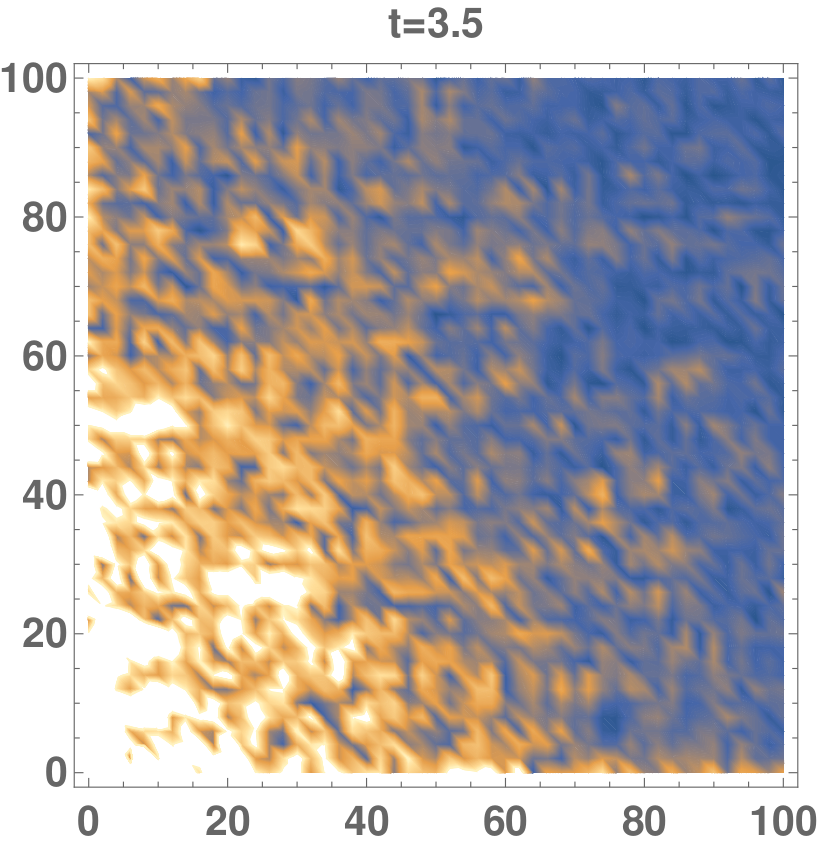}
      \includegraphics[width=0.05\linewidth]{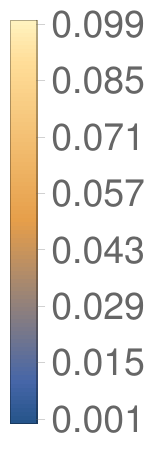}
	   \includegraphics[width=0.245\linewidth]{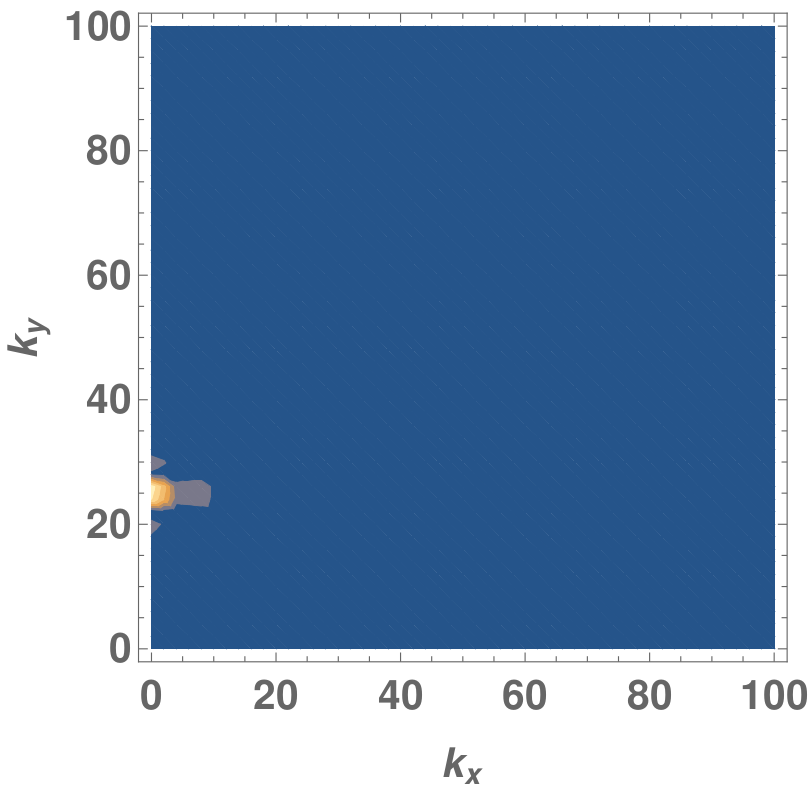}
        \includegraphics[width=0.225\linewidth]{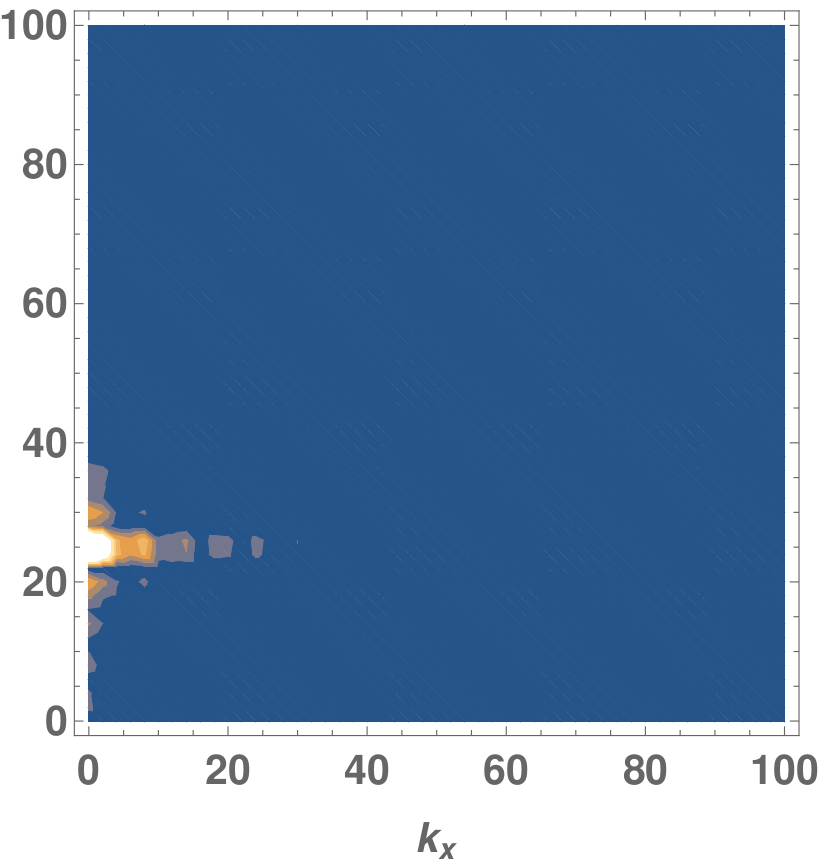}
         \includegraphics[width=0.225\linewidth]{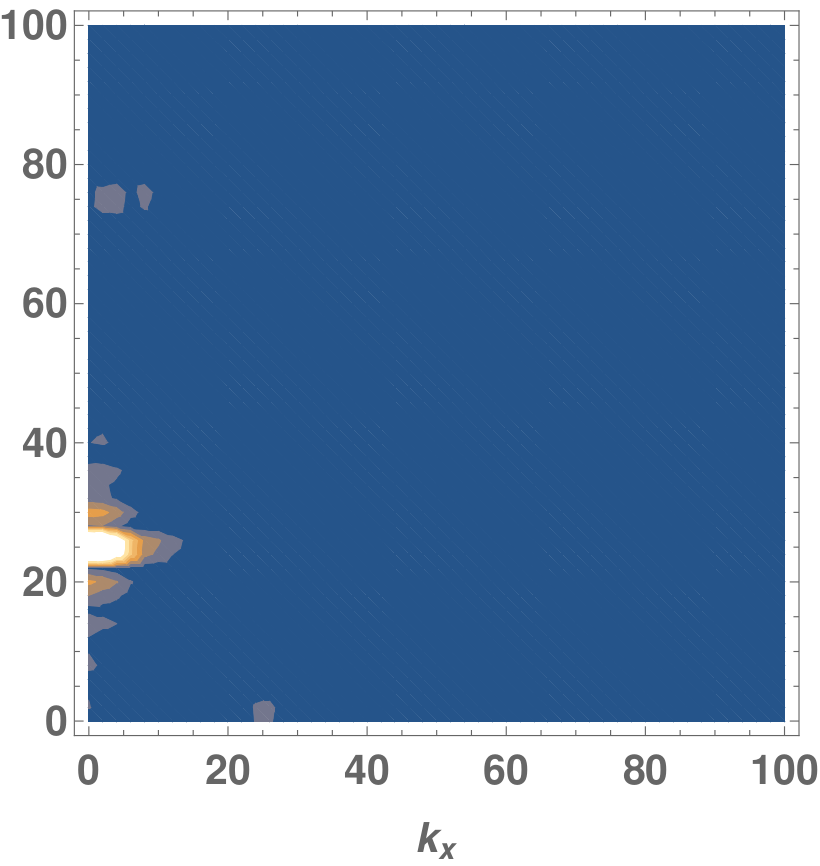}
          \includegraphics[width=0.225\linewidth]{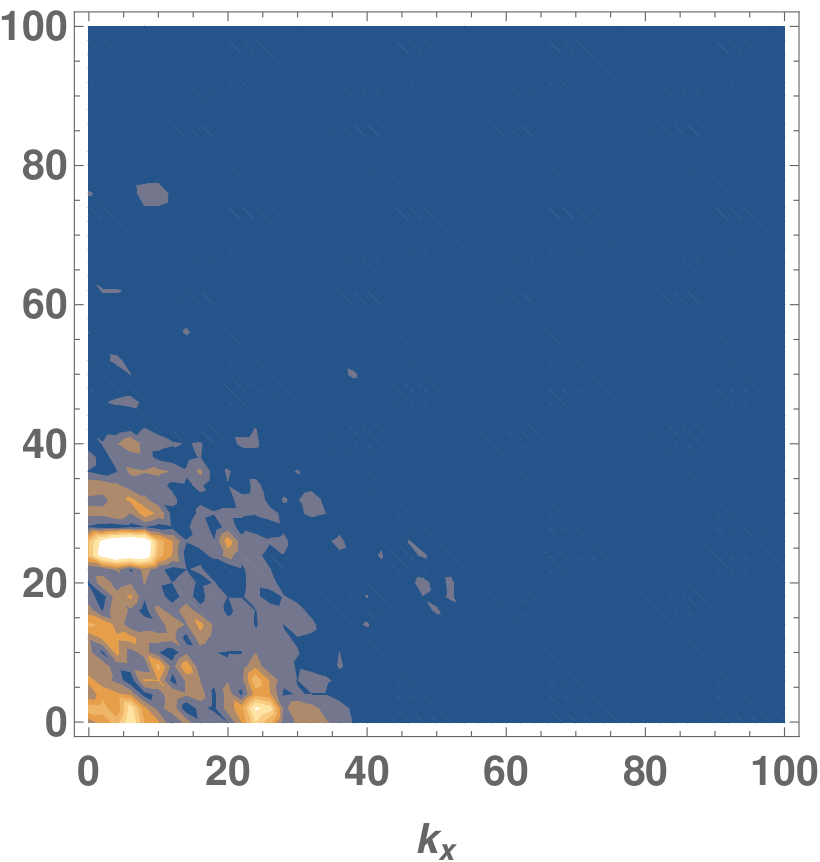}
           \includegraphics[width=0.05\linewidth]{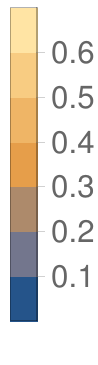}
  \caption{Figure shows the profile of modulus of field-modes $\lvert \tilde{\phi}_1(\vec{k},t)\lvert$ (upper panel) and 
  $\lvert \tilde{\phi}_2(\vec{k},t)\lvert$ (lower panel) at different times for $\omega$=50 $\Lambda^{-1}$, $\varepsilon$=0.4, 
  $\Phi_0$=10 $\Lambda^{-1}$, $\lambda$=40. From left to right, modes are plotted at times $t$=2.0, 2.5, 3.0, 3.5 $\Lambda$. 
  There is noticeable growth of $\lvert \tilde{\phi}_1(\vec{k},t)\lvert$ at $k_x$=0, $k_y$=$\omega$, and growth of 
  $\lvert \tilde{\phi}_2(\vec{k},t)\lvert$ at $k_x$=0, $k_y$=$\omega/2$. In the intermediate stage of field evolution, there 
  is also growth of $\lvert \tilde{\phi}_1(\vec{k},t)\lvert$ at $k_x$$\approx$$\omega/2$, $k_y$$\approx$$\omega/2$ and 
  $k_x$$\approx$$\omega/2$, $k_y$$\approx$$3\omega/2$. The subsequent evolution also leads to the growth of various other 
  momentum-modes of both fields.} 
  \label{fig5}
\end{figure*}

To see how fluctuations in fields $\phi_1$ and $\phi_2$ grow with time, we calculate field's fluctuations 
$\delta \phi_i (t)$=$\sqrt{\langle \phi_i^2(\vec{x},t)\rangle - \langle\phi_i(\vec{x},t) \rangle^2}$, where 
$i$=1,2, and bracket $\langle...\rangle$ represents the volume average (areal average in two dimensions) of 
the field. Fig.\ref{fig3} shows the time evolution of $\delta \phi_1$ and $\delta \phi_2$ for two sets of 
parameters of the effective potential. The parameters of spacetime metric for the simulations are 
$\varepsilon$=0.4 and $\omega$=100 $\Lambda^{-1}$, same as used in Figs.\ref{fig1},\ref{fig2}. Plots (a)(blue) 
and (b)(gray) are the time evolution of $\delta \phi_1$ and $\delta \phi_2$ for $\Phi_0$=10 $\Lambda^{-1}$ 
and $\lambda$=40, while plots (c)(black) and (d)(red) are the time evolution of $\delta \phi_1$ and 
$\delta \phi_2$ for $\Phi_0$=0.1 $\Lambda^{-1}$ and $\lambda$=4, respectively. Note that $\phi_1$ and 
$\phi_2$ at initial time are proportional to $\Phi_0$, which is the reason $\delta \phi_1$ and $\delta \phi_2$ 
start with larger values in plots (a) and (b) in comparison with plots (c) and (d) even though the chosen 
value of $\beta$ for both the sets of parameters of the effective potential is the same (indeed, the 
fluctuations in phase of field $\Phi$ is the same for both the sets of parameters of the effective potential). 
Fig.\ref{fig3} clearly shows that in each case, the growth of fluctuation in $\phi_2$ field is larger than fluctuation 
in $\phi_1$, which is again due to our choice of initial field configuration. After some time, both the field 
fluctuations become degenerate. Note that there are oscillations in plots (a) and (b) during the time interval 
of 0.5 - 1.5 $\Lambda$. These oscillations correspond to the large amplitude {\it field oscillations} as 
discussed in Fig.\ref{fig1}.

We now discuss a very crucial issue regarding the properties of formed vortices at different frequency regimes 
of spacetime oscillations. For two sets of parameters of the effective potential in Fig.\ref{fig3}, the masses of 
the field are $m_\Phi$=89.44 $\Lambda^{-1}$ and $m_\Phi$=0.28 $\Lambda^{-1}$, respectively. Therefore with the 
used frequency $\omega$ in this figure, for former case $m_{\Phi}$$<$$\omega$$<$$2m_{\Phi}$, while for latter case 
$\omega$$\gg$$m_{\Phi}$. Even though in both the cases there are qualitative similar growths in the fluctuations, 
there is a basic difference in the evolution of the field for these two cases. In the case of $\omega$$\gg$$m_{\Phi}$, 
spacetime oscillations generate waves in the field configuration having much shorter wavelength in comparison with 
$m_{\Phi}^{-1}$ (spacetime oscillations generate transverse-modes with momentum $\approx\omega/2$; see details later). 
Furthermore, later show that for high frequency of spacetime oscillations, the longitudinal-modes of the 
field also grow along with transverse-modes following the relation 
$|\vec{k}|^2$+$m_{\Phi}^2$=$\big(\frac{n\omega}{2}\big)^2$, where $|\vec{k}|$ is the momentum of the 
longitudinal-modes of the field, and $n$=1,2,3,.... Therefore in the case of $\omega$$\gg$$m_{\Phi}$, the fastest 
growing longitudinal-modes (modes for $n$=1 and $n$=2) must have much larger momentum in comparison with $m_{\Phi}$. 
All this leads to huge excitations in the field for $\omega$$\gg$$m_{\Phi}$, and the field acquires value much larger 
than VEV in a short time. In such a situation, the field does not feel the presence of central barrier of effective 
potential and does go back and forth across this easily. As the core size of a topological vortex is given by 
$m_{\Phi}^{-1}$ where the field magnitude at distances larger than $m_{\Phi}^{-1}$ is given by VEV, therefore under 
such a huge field excitations having much shorter wavelength in comparison with $m_{\Phi}^{-1}$, there is no  
possibility of the formation of topological vortices. Thus, in the case of $\omega$$\gg$$m_{\Phi}$, topological vortices 
cannot be formed under the spacetime oscillations. Note that for second set of parameters of the effective potential, 
our lattice size is smaller than $m_{\Phi}^{-1}$, hence it cannot accommodate even a single topological vortex for 
these parameters. However, our above conclusion holds true even for a lattice whose size is much bigger than 
$m_{\Phi}^{-1}$.  

However, in this case with sufficient evolution of the field, after time $t$$\simeq$2.7 $\Lambda$, we see the 
formation of some vortex-antivortex kind of structures (even on the used lattice), though these are not topological 
vortices. In these vortices, due to huge excitations, the field acquires values way beyond the VEV outside the vortex 
core; note that in this case, VEV is 0.1 $\Lambda^{-1}$, where outside these vortices, at this time, field acquires 
value $\simeq$20 $\Lambda^{-1}$, which is 200 times larger than VEV. Furthermore, in this case, we have found that the 
vortex core size is given by $\omega^{-1}$ (instead of $m_{\Phi}^{-1}$), as expected from the wavelength 
of the generated field excitations. Thus, these vortices certainly do not satisfy the properties of topological vortices, 
rather their properties are determined by the parameters of spacetime oscillations. Basically in this case, the magnitude 
of the field becomes so large in comparison with VEV that field configuration is able to create vortices (non-topological) 
with core size of much shorter length scale than $m_{\Phi}^{-1}$. (We show the vortices in the case of 
$\omega$$\gg$$m_{\Phi}$ in Fig.\ref{fig14}, where $m_{\Phi}$=0.28 $\Lambda^{-1}$ and $\omega$=17 $\Lambda^{-1}$ have 
been used for the simulation).

On the other hand, in the former case as discussed above, although longitudinal excitation arises and hence vortices are 
not well formed, vortices up to some extent satisfy the properties of topological vortices; see Fig.\ref{fig1} and 
Fig.\ref{fig2}. Later, we show that there is a fundamental difference in the evolution of field for 
$\omega$$>$$m_\Phi$ and for $\omega$$\leq$$m_\Phi$. In the case of $\omega$$\leq$$m_{\Phi}$, the field remains close to 
VEV where mainly transverse excitation arises under the spacetime oscillations. Therefore, the formed vortices in such 
case certainly are topological vortices; see Fig.\ref{fig10}. Thus, only in those cases, for which frequency is either 
less than or close to the mass of the field, i.e. for which $\omega$$\lesssim$$m_{\Phi}$, the formed vortices under 
the spacetime oscillations are topological in nature. In our simulations, however, the field-profile of these vortices 
never achieves an exact profile of topological vortices as the field configuration has to bear continuous spacetime 
oscillations.    

To further investigate the behavior of field configurations at various time stages, we calculate the time evolution of   
various momentum-modes of the fields $\phi_1$ and $\phi_2$. The Fourier transform of the field configuration from the physical 
space to the momentum space at any time is given by,  
\begin{equation}
  \tilde{\phi}_i(\vec{k},t) = \frac{1}{\mathcal{A}}\int_{b.c.}d^2\vec{x}~ \phi_i(\vec{x},t) e^{i\vec{k}.\vec{x}}~; ~~i=1,2,
\end{equation}
where $\mathcal{A}$ is the total area of the system, {\it b.c.} stands for the {\it boundary condition}, and 
$\vec{k}$=$k_x \hat{x}$+$k_y \hat{y}$, $\vec{x}$=$x \hat{x}$+$y \hat{y}$ are the momentum and position vectors, respectively. 
In Fig.\ref{fig4} and Fig.\ref{fig5}, we plot modulus of momentum-modes of fields $\phi_1$ and $\phi_2$, i.e. 
$\lvert \tilde{\phi}_1(\vec{k},t)\lvert$ (upper panel) and $\lvert \tilde{\phi}_2(\vec{k},t)\lvert$ (lower panel), in the 
momentum space with spacing $\Delta k_x$=$\Delta k_y$=2.0 $\Lambda^{-1}$. (The white regions in plots correspond to the 
magnitude larger than the maximum range of legend.) In Fig.\ref{fig4}, we have taken spacetime oscillation frequency larger 
than $m_\Phi$, while in Fig.\ref{fig5}, the frequency is taken smaller than that. Therefore the field dynamics is expected to be
different in both the cases. Note that for $\phi_1$ field, zero-mode (zero-momentum mode) is the most dominant mode, while 
for $\phi_2$, it is close to zero. This is just because of initially chosen values of these fields. In Fig.\ref{fig4}, we have 
taken $\omega$=100 $\Lambda^{-1}$, $\varepsilon$=0.4, $\Phi_0$=10 $\Lambda^{-1}$, and $\lambda$=40 for the simulation. From left to 
right, field-modes are plotted at times $t$=0.05, 1.05, 1.35, 1.8 $\Lambda$; see also Fig.\ref{fig1} and Fig.\ref{fig2}. Note 
that the time interval 0.5 - 1.5 $\Lambda$ is the time-stage of large amplitude field oscillations for the given 
$\omega$. At the initial time itself, all higher momentum-modes of the field up to $k_x$=$k_y$=$2\pi/\Delta x$ are present 
due to the prescribed random fluctuations, which can be seen in the plot at time $t$=0.05 $\Lambda$. Under the spacetime 
oscillations, depending upon the frequency $\omega$, some specific momentum-modes of field grow with time. In the figure, 
there is noticeable growth of $\lvert \tilde{\phi}_1(\vec{k},t)\lvert$ at $k_x$=15 $\Lambda^{-1}$, $k_y$=$\omega$ and 
$k_x$$\approx$$k_y$$\approx$$\sqrt{\big(\omega^2-(\frac{m_\Phi}{\sqrt{2}})^2\big)/2}$ =54.77 $\Lambda^{-1}$, and growth of
$\lvert \tilde{\phi}_2(\vec{k},t)\lvert$ at $k_x$=0, $k_y$=$\omega/2$ and $k_x$=$\omega/2$, $k_y$=0. The subsequent evolution 
also leads to the growth of various other momentum-modes of both fields.

In Fig.\ref{fig5}, we have taken $\omega$=50 $\Lambda^{-1}$, $\varepsilon$=0.4, $\Phi_0$=10 $\Lambda^{-1}$, and $\lambda$=40 
for the simulation. From left to right, field-modes are plotted at times $t$=2.0, 2.5, 3.0, 3.5 $\Lambda$. Note that for 
the given $\omega$, the time interval 2.0 - 3.7 $\Lambda$ corresponds to the stage of large amplitude oscillations of 
the field. In this case also, the initial distributions of field-modes are same as plotted in Fig.\ref{fig4} at time 
$t$=0.05 $\Lambda$ (therefore, not shown again in Fig.\ref{fig5}). Figure clearly shows that there is noticeable growth of 
$\lvert \tilde{\phi}_1(\vec{k},t)\lvert$ at $k_x$=0, $k_y$=$\omega$, and growth of $\lvert \tilde{\phi}_2(\vec{k},t)\lvert$ 
at $k_x$=0, $k_y$=$\omega/2$. In the intermediate stage of field evolution, there is also growth of 
$\lvert \tilde{\phi}_1(\vec{k},t)\lvert$ at $k_x$$\approx$$\omega/2$, $k_y$$\approx$$\omega/2$ and $k_x$$\approx$$\omega/2$, 
$k_y$$\approx$$3\omega/2$. In this case also, the subsequent evolution leads to the growth of various other momentum-modes 
of both fields. 

To understand why some specific momentum-modes of the field grow for a given $\omega$, we write the field as 
$\Phi$=$\Phi_0$+$v_1$+$iv_2$ ($\phi_1$=$\Phi_0$+$v_1$, $\phi_2$=$v_2$), where $v_1$ and $v_2$ are small fluctuations 
of the field about the VEV. We write the equation of motion for these fluctuations keeping terms up to linear order 
in these fields, and then Fourier transform in the momentum space. The evolution equations for momentum-modes of 
these fluctuations are 
\begin{equation}
  \frac{\varepsilon^2 \omega \sin (2\omega t)}{2~f(t)~f(-t)} 
  \dot{\tilde{v}}_{ik} + \ddot{\tilde{v}}_{ik} + \big(k_x^2 ~f(t) + k_y^2 ~f(-t) + m_i^2 \big)\tilde{v}_{ik}=0,
\label{eom2}
\end{equation}
where $\tilde{v}_{ik}$$\equiv$$\tilde{v}_i(\vec{k},t)$, $f(t)$=$1$$-$$\varepsilon\sin(\omega t)$, 
$m_1$=$m_\Phi$, and $m_2$=0. These equations clearly show that $\tilde{v}_{1k}$ and $\tilde{v}_{2k}$ undergo 
parametric resonance (get resonant growths) if suitable conditions (relations between $k_x$, $k_y$, $m_i$, 
and $\omega$) are satisfied. We numerically solve these equations and find that there are various resonance 
conditions for both fields. In a reasonable range of frequency $\omega$, the resonance conditions for 
$\tilde{v}_{ik}$ are $|\vec{k}|^2$+$m_i^2$=$\big(\frac{n\omega}{2}\big)^2$, where $n$=1,2,3,... . For each 
$n$, there are different frequency cut-off for $\tilde{v}_{1k}$ to get resonance growth, which are 
$\omega$$\leq$$2m_{\Phi}/n$. This indicates that the most dominant modes of $\tilde{v}_{1k}$, i.e. modes for 
$n$=1 and $n$=2, grow only when $\omega$ is greater than $m_{\Phi}$. Thus, effectively the frequency cut-off 
for the longitudinal-modes is $\omega$$\leq$$m_{\Phi}$. On the other hand, for $\tilde{v}_{2k}$, there is no 
such frequency 
cut-off (due to its zero mass), therefore it can undergo parametric resonance even for any low frequency 
$\omega$ (where $\omega$$>$0). For a comparison, if one ignores the first term from the equation of 
$\tilde{v}_{2k}$, then in (1+1)-dimensions it takes the form of a {\it harmonic oscillator equation} with the 
(time dependent) frequency square $k_{x(y)}^2\big(1$$\mp$$\sin(\omega t)\big)$ \cite{landauMech}. In such case, 
the oscillator undergoes {\it parametric resonance} when the condition $k_{x(y)}$$\simeq$$n\omega/2$ is satisfied 
\cite{landauMech}. For a given frequency $\omega$, the growth rates of growing modes of $\tilde{v}_{ik}$ are 
different, which get changed by changing $\omega$. The fastest growing modes out of $\tilde{v}_{1k}$ and 
$\tilde{v}_{2k}$ are $\tilde{v}_{2k}$ at $k_{x(y)}$=$\omega/2$, $k_{y(x)}$=0, whose growth rates decrease by 
decreasing frequency of spacetime oscillations. This response of field-modes to the frequency $\omega$ can play 
an important role in the understanding of the time of formation of first vortex-antivortex pair in the system at 
different frequencies $\omega$. 

A similar feature of the resonance growths as discussed above can be seen in Fig.\ref{fig4} and Fig.\ref{fig5}, 
where $\tilde{\phi}_2$ grows resonantly at $k_{x(y)}$$\simeq$$\omega/2$, $k_{y(x)}$$\simeq$0. In these figures, 
there is a qualitative difference in the evolution of $\tilde{\phi}_1$ for $\omega$$>$$m_\Phi$ and 
$\omega$$<$$m_\Phi$, as for $\omega$$>$$m_\Phi$ some momentum-modes of the longitudinal component ($\tilde{v}_{1k}$) 
also grow resonantly. This shows that in the frequency regime of $\omega$$>$$m_\Phi$, both components of the 
field [transverse as well as longitudinal] are generated under the resonance process, whereas in the frequency 
regime of $\omega$$\leq$$m_\Phi$, a predominantly transverse component is generated. It has to be noted that in 
Fig.\ref{fig4}, 
$\tilde{\phi}_1$ grows by following relation $|\vec{k}|^2$+$(\frac{m_\Phi}{\sqrt{2}})^2$=$\omega^2$ instead of 
following relation obtained for $\tilde{v}_{1k}$ at $n$=2. This is happening probably because $\phi_1$ is the 
field whose values are given from the center of the effective potential. It is important to appreciate that the 
appropriate momentum-modes of the field corresponding to $\omega$ have to be present initially to have resonance 
growths, where transverse-modes of the field can grow even at frequencies less than the mass of the field. Thus, 
under the spacetime oscillations, there is no lowest frequency cut-off for resonance to occur as long as the 
appropriate momentum-modes of the field are present initially to grow (however, finite size effects of the system 
set a lowest frequency cut-off; see discussions later). We point out that the above discussion is based on the 
linear approximation, while $\phi_1$ and $\phi_2$ are the coupled fields whose dynamics are modulated by the 
effective potential. This couples various momentum-modes of the field non-linearly, and makes resonance conditions 
much more complicated. 

\begin{figure}
  \includegraphics[width=1.\linewidth]{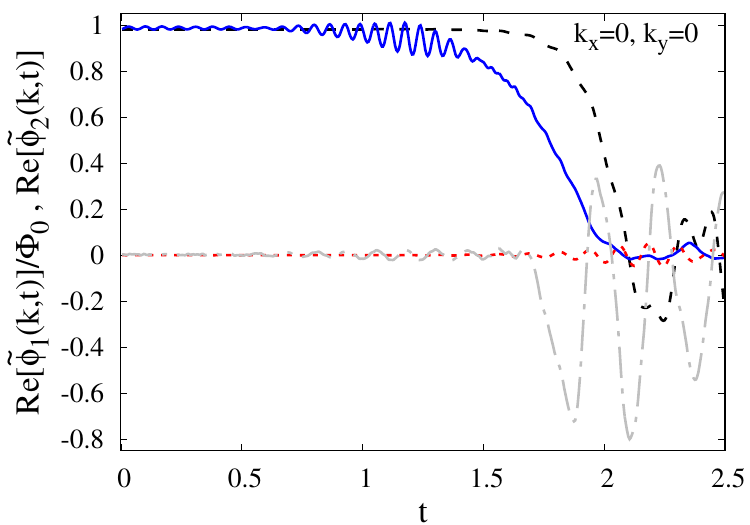}
  \caption{Figure shows the time evolution of real part of zero-modes of fields $\phi_1$ and $\phi_2$ for two sets of parameters 
  of effective potential. In the figure, zero-mode of $\phi_1$ field is normalized with the respective $\Phi_0$ value. We have 
  taken $\varepsilon$=0.4 and $\omega$=100 $\Lambda^{-1}$ for the simulation. Solid(blue) and dash-dotted(gray) lines are the 
  time evolution of zero-modes of fields $\phi_1$ and $\phi_2$ for $\Phi_0$=10 $\Lambda^{-1}$ and $\lambda$=40, while dashed(black) 
  and dotted(red) lines are for $\Phi_0$=0.1 $\Lambda^{-1}$ and $\lambda$=4, respectively. Note that there are small amplitude 
  oscillations in dashed(black) line, which does not appear because of large $y$-range.} 
  \label{fig6}
\end{figure}

The above analysis shows that the higher momentum-modes of fields $\phi_1$ and $\phi_2$, following their respective 
resonance conditions, grow with time. Now we show how zero-modes (modes for $k_x$=0, $k_y$=0) of these fields vary 
with time. The zero-modes of fields characterise the field configuration at each time and provide the frequency of 
oscillations during the large amplitude oscillation stage of the field. Fig.\ref{fig6} shows the time evolution of 
real part of zero-modes of fields $\phi_1$ and $\phi_2$ for two sets of parameters of the effective potential. Note 
that the imaginary part of zero-modes for both the fields are zero. In the figure, zero-mode of $\phi_1$ field is 
normalized with the respective $\Phi_0$ value. For this simulation, we have taken $\varepsilon$=0.4 and $\omega$=100 
$\Lambda^{-1}$. Solid(blue) and dash-dotted(gray) lines are the time evolution of zero-modes of fields $\phi_1$ and 
$\phi_2$, respectively, for $\Phi_0$=10 $\Lambda^{-1}$, $\lambda$=40. In this case, there are oscillations in the 
$\phi_1$ zero-mode during the time interval of 0.5 - 1.5 $\Lambda$ with growing and then decaying amplitude; let us 
call it the {\it large amplitude field oscillation} stage. Roughly at time $t$=1.5 $\Lambda$, these oscillations stop 
and the zero-mode starts collapsing towards the zero value. The collapse of zero-mode starts roughly at the time of 
formation of vortex-antivortex pairs in the system. We have performed the simulations for different spacetime 
oscillation frequencies and calculated the time evolution of $\phi_1$ zero-mode for each case (in the figure, it is 
not shown for other frequencies). We see that in all cases, at initial time there is a unique frequency and amplitude 
of oscillations of $\phi_1$ zero-mode, which with subsequent evolution get modulated with a {\it new} frequency of 
oscillations with growing amplitude as shown in Fig.\ref{fig6} from time $t$=0.5 $\Lambda$. This shows that the initial 
time oscillations of $\phi_1$ zero-mode (before time $t$=0.5 $\Lambda$ in Fig.\ref{fig6}) are independent from the 
spacetime oscillations, and may be related with the evolution of initial field fluctuations. On the other hand, the 
{\it new} modulation frequency, i.e. the frequency during the time-stage of large amplitude oscillations of $\phi_1$ 
zero-mode, is exactly equal to the spacetime oscillation frequency $\omega$ (which we have shown later). This gives  
more evidence of the {\it parametric resonance} of the field under the spacetime oscillations. 

\begin{figure}
  \includegraphics[width=1.\linewidth]{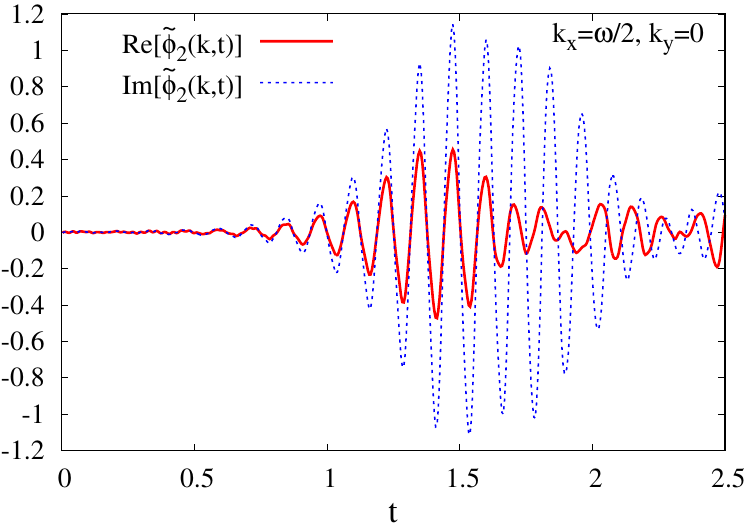}
  \caption{Figure shows the time evolution of real and imaginary part of ($k_x$=$\omega/2$, $k_y$=0)-mode of field $\phi_2$ with 
  solid(red) and dotted(blue) lines, respectively, for $\omega$=100 $\Lambda^{-1}$, $\varepsilon$=0.4, $\Phi_0$=10 $\Lambda^{-1}$, 
  $\lambda$=40.} 
  \label{fig7}
\end{figure}

In Fig.\ref{fig6}, zero-mode of $\phi_2$ field (shown by dash-dotted(gray) line) also has small amplitude oscillations roughly in 
the same time interval (i.e. in time interval of 0.5 - 1.5 $\Lambda$), but without much growth. Later, we show that the
oscillation frequency of zero-mode of $\phi_2$ during this time interval is half the spacetime oscillation frequency. Here, it 
should be noted that during this time interval, $\phi_2$ varies in between negative and positive values spatially as well as
temporarily, as can be seen in Fig.\ref{fig1}(b). Therefore even though the absolute value of $\phi_2$ reaches a non-zero value 
in some regions, its zero-mode comes out to be very small (just because of the cancellation of positive and negative values). 
This is the reason why the zero-mode of $\phi_2$ does not show any growth during this time interval (indeed, this mode does not 
follow the resonance condition). It is already clear from Fig.\ref{fig3} that during this time interval, the growth in the
fluctuations of field $\phi_2$ is larger in comparison with fluctuations in $\phi_1$. Therefore, the higher momentum-modes of 
$\phi_2$ must show growths with time, which already has been seen in Fig.\ref{fig4} at $k_x$=$\omega/2$, $k_y$=0 and $k_x$=0, 
$k_y$=$\omega/2$. In Fig.\ref{fig7}, we plot the time evolution of the real and imaginary parts of ($k_x$=$\omega/2$, 
$k_y$=0)-mode of $\phi_2$ with solid(red) and dotted(blue) lines, respectively. This clearly shows oscillations of these modes 
with growing and then decaying amplitude with time.  

To show the dependence of the zero-modes of fields $\phi_1$ and $\phi_2$ on the parameters of the effective potential, 
in Fig.\ref{fig6}
we also plot their time evolution for $\Phi_0$=0.1 $\Lambda^{-1}$ and $\lambda$=4 with dashed(black) and dotted(red) lines, 
respectively. In this case, depth of the effective potential is very small, therefore the field can easily climb the central
barrier at $\Phi$=0. Note that in this case, the amplitude of oscillations of $\phi_1$ zero-mode is small, which is not clearly 
visible because of the large $y$-range in the plot. The amplitude of oscillations of $\phi_2$ zero-mode is also very small in the 
plot because of the same reason mentioned above.

\begin{figure}
  \includegraphics[width=1.\linewidth]{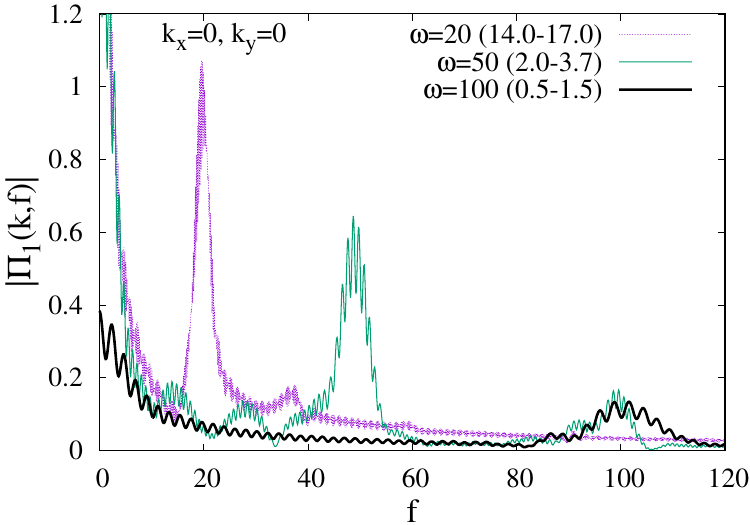}
  \caption{Figure shows the frequency spectrum of $\phi_1$ zero-mode for different spacetime oscillation frequencies $\omega$ in the time 
  interval in which it oscillates with a significant large amplitude. This time interval for the respective frequencies is written in 
  brackets. Violet(dotted), green(solid thin), and black(solid thick) lines correspond to the frequency spectrum for $\omega$=20, 
  50, and 100 $\Lambda^{-1}$, respectively. In each case, the frequency spectrum has a peak at the respective frequency 
  $\omega$ of spacetime oscillations, which indicates the phenomenon of {\it parametric resonance} of the field.}
  \label{fig8}
\end{figure}

To determine the oscillation frequency of fields during the time interval of $t_i$ - $t_f$ in which $\phi_1$ zero-mode 
oscillates with a significantly large amplitude, we perform Fourier transform of fields' zero-mode from $t$-space to the frequency 
$f$-space as, 
\begin{equation}
  \Pi_i(\vec{k},f) = \frac{1}{(t_f-t_i)}\int_{t_i}^{t_f} dt~ \tilde{\phi}_i(\vec{k},t) e^{-ift}~; ~~i=1,2.
\end{equation}
We calculate the {\it moduli} of $\Pi_1(\vec{k},f)$ and $\Pi_2(\vec{k},f)$, which provide the frequency spectrum of momentum-modes 
of $\phi_1$ and $\phi_2$ fields. Fig.\ref{fig8} shows the frequency spectrum of zero-mode of $\phi_1$ during the time interval of 
$t_i$ - $t_f$ for different spacetime oscillation frequencies $\omega$. In the figure, such a time interval for each frequency 
$\omega$ is mentioned in brackets. To obtain this frequency spectrum, in each case, $\Phi_0$ has been subtracted from the $\phi_1$ 
zero-mode so that the background frequency modes, arising due to constant $\Phi_0$ value, get eliminated and the peak structures 
become apparent. (However, this subtraction does not remove background frequency modes completely, which is the reason a peak at the 
zero frequency is still present in each plot). In the figure, violet(dotted), green(solid thin), and black(solid thick) lines 
correspond to the frequency spectrum for $\omega$=20, 50, and 100 $\Lambda^{-1}$, respectively. In 
each case, $\varepsilon$=0.4, $\Phi_0$=10 $\Lambda^{-1}$, and $\lambda$=40 have been taken for simulations. In the figure, in each 
case, the frequency spectrum has a peak at the respective frequency $\omega$ of spacetime oscillations, suggesting the phenomenon of 
{\it parametric resonance}. Note that, for the frequency $\omega$=50 $\Lambda^{-1}$, there are two peaks at 
$f$=50 $\Lambda^{-1}$(dominant peak) and at $f$=100 $\Lambda^{-1}$(sub-dominant peak). The second peak is arising may be due to 
coefficient $\sin(2\omega t)$ of the first order time derivative term in Eq.(\ref{eom}), or may be due to resonance process itself.
For $\Phi_0$=0.1 $\Lambda^{-1}$ and $\lambda$=4, the amplitude of oscillations of $\phi_1$ zero-mode is too small to determine such peak 
structure, that is why frequency spectrum for this case has not been shown in the figure.  

\begin{figure}
  \includegraphics[width=1.\linewidth]{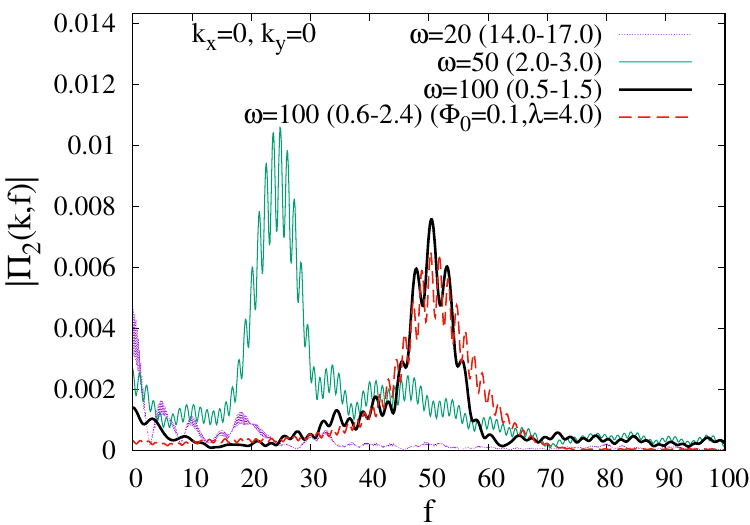}
  \caption{Figure shows the frequency spectrum of $\phi_2$ zero-mode for different spacetime oscillation frequencies $\omega$ in the same 
  time interval as mentioned in Fig.\ref{fig8} (mentioned here as well in brackets). Violet(dotted), green(solid thin), and black(solid 
  thick) lines correspond to the frequency spectrum for $\omega$=20, 50, and 100 $\Lambda^{-1}$, respectively. 
  The peak structures show that the zero-mode of $\phi_2$ oscillates with frequency $\omega/2$. Red(dashed) line corresponds to the case for 
  $\omega$=100 $\Lambda^{-1}$, $\Phi_0$=0.1 $\Lambda^{-1}$, and $\lambda$=4, which also has a peak at $\omega/2$.}  
  \label{fig9}
\end{figure}

Fig.\ref{fig9} shows the frequency spectrum of zero-mode of $\phi_2$ in the same time interval as mentioned in Fig.\ref{fig8} for 
different spacetime oscillation frequencies $\omega$ (the time interval $t_i$ - $t_f$ for each frequency $\omega$ is mentioned in 
brackets). Unlike for $\phi_1$ zero-mode, in this case, there is no background frequency spectrum. In the figure, violet(dotted), green(solid 
thin), and black(solid thick) lines correspond to the frequency spectrum for $\omega$=20, 50, and 100 
$\Lambda^{-1}$, respectively. In this case, each frequency spectrum has peak at half the spacetime oscillation frequency $\omega$. For 
$\omega$=20 $\Lambda^{-1}$, there is no unique and strong peak structure in the frequency spectrum. In the figure, red(dashed) line 
corresponds to the case for $\omega$=100 $\Lambda^{-1}$, $\Phi_0$=0.1 $\Lambda^{-1}$, and $\lambda$=4.0. In this case also, frequency 
spectrum has peak at the half $\omega$, which shows the independence of some aspects of this phenomenon from the parameters of the 
effective potential. We have also checked that the real and imaginary parts of momentum-modes of $\phi_2$ at $k_x$=$\omega/2$, $k_y$=0 
and $k_x$=0, $k_y$=$\omega/2$ (see Fig.\ref{fig7}) also oscillate with frequency $\omega/2$. 

With all these results, it is clear that under the spacetime oscillations, the field $\Phi$ itself starts performing oscillations in space 
and time with certain momentum and frequency by following the resonance condition for the given $\omega$, where $\phi_1$ field oscillates with 
frequency $\omega$ and $\phi_2$ oscillates with frequency $\omega/2$. This shows that the field $\Phi$ undergoes {\it parametric resonance} 
under the spacetime oscillations.

\subsection{B. Frequency dependence of generation of field excitations}

As discussed earlier, to generate longitudinal excitation dominantly, the frequency of spacetime oscillations should be 
$\omega$$>$$m_\Phi$. In contrary to this, to generate transverse excitation of the field, there is no such frequency cut-off as 
the mass of modes corresponding to this excitation is zero. Therefore even in the low frequency regime, i.e. $\omega$$<$$m_\Phi$,
spacetime oscillations can lead to the generation of transverse excitation via parametric resonance. Thus, there is a fundamental 
difference in the evolution of field in the frequency regimes $\omega$$>$$m_\Phi$ and $\omega$$\leq$$m_\Phi$, where in both the 
regimes, in principle, the field excitations can be generated under the spacetime oscillations.

\begin{figure}
 \includegraphics[width=0.475\linewidth, angle=270]{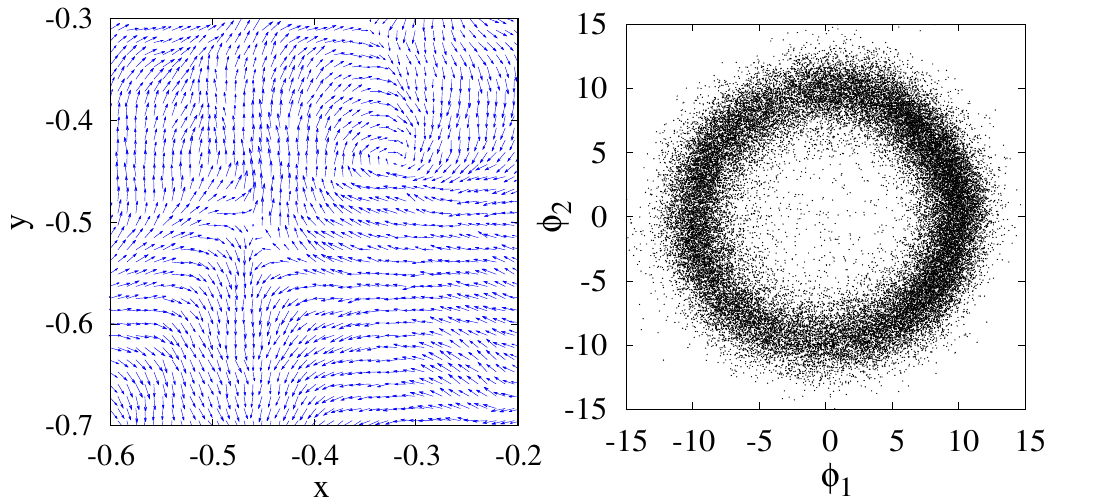}
\caption{Figure shows the field configuration in physical space (left panel) and in field-space (right panel) at time $t$=18.5 $\Lambda$ for 
the frequency $\omega$=20 $\Lambda^{-1}$ ($\varepsilon$=0.4, $\Phi_0$=10 $\Lambda^{-1}$, and $\lambda$=40). Left plot shows the formation of 
well separated vortex-antivortex pair in the system. Right plot shows that at this frequency of spacetime oscillations, mainly transverse 
excitation of the field has been generated.}
 \label{fig10}
\end{figure}

In our simulation, we have studied the generation of field excitations and the formation of vortex-antivortex pairs for a wide range 
of frequencies of spacetime oscillations. At low frequencies, vortex-antivortex pairs form with relatively smaller vortex densities 
and remain well separated. In this frequency regime, unlike the case of high frequency, field remains close to the VEV of effective 
potential for significantly longer time of field evolution. This is shown in Fig.\ref{fig10}, where left panel shows the vector 
plot of field configuration in the physical space and right panel shows the field distribution in the field-space; both are at time 
$t$=18.5 $\Lambda$. For this simulation, the spacetime oscillation frequency is taken as $\omega$=20 $\Lambda^{-1}$, and other parameters 
of simulation are $\varepsilon$=0.4, $\Phi_0$=10 $\Lambda^{-1}$, and $\lambda$=40 (same as used earlier). Left plot clearly shows the 
formation of well separated vortex-antivortex pair in the system. Right plot shows that at almost all lattice points, the field is 
close to VEV of the effective potential, where the field distribution covers the whole vacuum manifold at this time (at initial time, 
the field distribution is a localized distribution in the field-space; see Fig.\ref{fig2}(a)). It shows that mainly transverse 
excitation has been generated in the system, while longitudinal excitation is strongly suppressed in this frequency regime; compare 
with Fig.\ref{fig2}(d).

In extremely low frequency regime of spacetime oscillations ($\omega$$\lesssim$$4\pi$(system~size)$^{-1}$; see discussion below), finite 
size of the system affects the generation of field excitations. This happens because spacetime oscillations couple to $k$-modes of the 
field, where for $\phi_2$ field, ($k_{x(y)}$=$\omega/2$, $k_{y(x)}$=0)-modes dominantly grow with time; see Fig.\ref{fig4} and Fig.\ref{fig5}. 
For a finite lattice, the $k$-mode with a wavelength equal to the system size $L$ will be $k_{_L}$=$2\pi/L$.  
Therefore any frequency $\omega$ below the cut-off $2 k_{_L}$ cannot lead to the growth of the most dominant resonance modes of the field; 
excitations with wavelength larger than the lattice size cannot grow resonantly. Therefore to excite the field, the frequency of spacetime 
oscillations must be $\omega$$\geq$$4\pi/L$ or equivalently $f_a$$\geq$$2/L$, where $f_a$=$\omega/2\pi$. In our simulations, even with 
frequencies two or three times larger than this cut-off, the field finds difficulties in achieving excitations. We have checked that, the
field excitations, which cannot be generated on a small lattice with a given spacetime oscillation frequency, can be generated using a larger 
lattice with the same frequency. This suggests that, in principle, any low frequency of spacetime oscillations can excite the field leading 
to the formation of vortices if a suitably large system size is chosen and the field is evolved for a significantly long time; see details in 
the next subsection.

\subsection{C. Time dependence of formation of vortices on parameters of spacetime oscillations}

In this subsection, we present our study on the time dependence of formation of vortices on parameters of spacetime oscillations. For this, 
we note down time at which first vortex-antivortex pair forms in the system. We call this time as the {\it vortex formation time} and denote 
it by $t_{vortex}$. To locate the vortices, we compute the variation of phase of $\Phi$ around each elementary square of lattice. The variation 
of phase between any two adjacent points on the elementary square is determined by the {\it geodesic rule} \cite{geodesic}. By continuity, 
the phase of a complex scalar field along a loop is always $2\pi n$, where $n$ is an integer. Throughout the evolution of field, we observe 
only $n$=$0,\pm 1$ for the variation of phase around each elementary square, where we identify vortex with $n$=+1 and antivortex with $n$=$-1$. 
In our simulations, we observe that $t_{vortex}$ does not depend much on the amplitude of fluctuations (value of $\beta$) of the initial field
configuration.

\begin{figure}
 \includegraphics[width=1.0\linewidth]{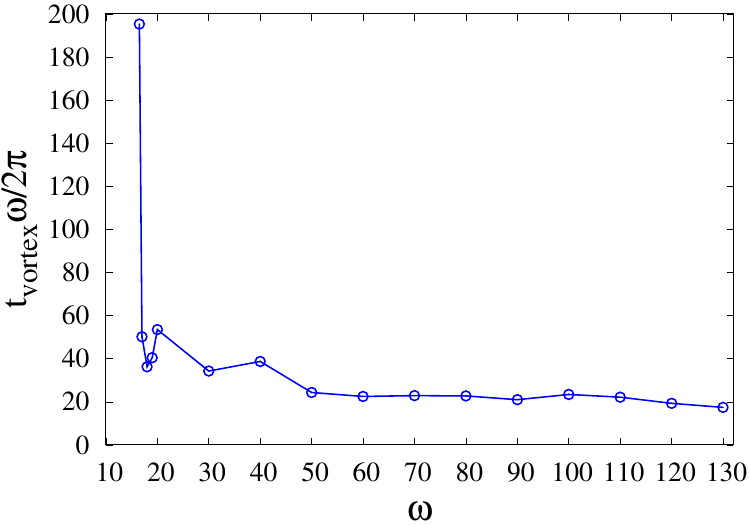}
 \caption{Figure shows the dependence of vortex formation time $t_{vortex}$ on various frequencies $\omega$. In figure, we have plotted the 
 {\it number of cycle} of spacetime oscillations up to the time $t_{vortex}$, i.e. $t_{vortex}~\omega/2\pi$, versus $\omega$.  
 Other parameters of simulation are $\varepsilon$=0.4, $\Phi_0$=10 $\Lambda^{-1}$, and $\lambda$=40. Figure clearly shows that in the 
 frequency regime $\omega$$\gtrsim$50 $\Lambda^{-1}$, $t_{vortex}~\omega/2\pi$ is almost independent from $\omega$. However, by decreasing 
 frequency below $\omega$=50 $\Lambda^{-1}$, this number starts increasing.}
 \label{fig11}
\end{figure}

In Fig.\ref{fig11}, we plot $t_{vortex}~\omega/2\pi$ versus $\omega$ for $\varepsilon$=0.4. The parameters of the effective potential 
are $\Phi_0$=10 $\Lambda^{-1}$ and $\lambda$=40. Note that $t_{vortex}~\omega/2\pi$ is a dimensionless number which is equal to the {\it number 
of cycle} of spacetime oscillations up to the time $t_{vortex}$. Figure clearly shows that in the frequency regime $\omega$$\gtrsim$50 
$\Lambda^{-1}$, the {\it number of cycle} required to form vortices is almost independent from $\omega$. However, by decreasing the frequency 
of spacetime oscillations below $\omega$=50 $\Lambda^{-1}$, this number starts increasing. We have not seen the formation of any vortices below 
$\omega$=16 $\Lambda^{-1}$ up to a significantly long time of simulation on the used lattice. 

As mentioned earlier, in extremely low frequency regime, finite size of the system restricts the generation of field excitations, 
therefore it must also restrict the formation of vortices. In our simulation, we observe that up to a significantly long time, 
there is no formation of any vortex-antivortex pair on the 200$\times$200 lattice with frequency $\omega$=15 $\Lambda^{-1}$, however 
with the same parameters of simulation, vortices form on the 800$\times$800 lattice. Therefore, it is quite possible that in Fig.\ref{fig11}, 
the deviations of $t_{vortex}~\omega/2\pi$ below $\omega$=50 $\Lambda^{-1}$ from a constant value have arisen due to the finite size 
effects of the system. Therefore, it may also be possible that this constant value of $t_{vortex}~\omega/2\pi$ is a universal number 
of cycle for formation of vortices, and has the same value even in the low frequency regime when finite size effects are eliminated. 

\begin{figure}
 \includegraphics[width=1.0\linewidth]{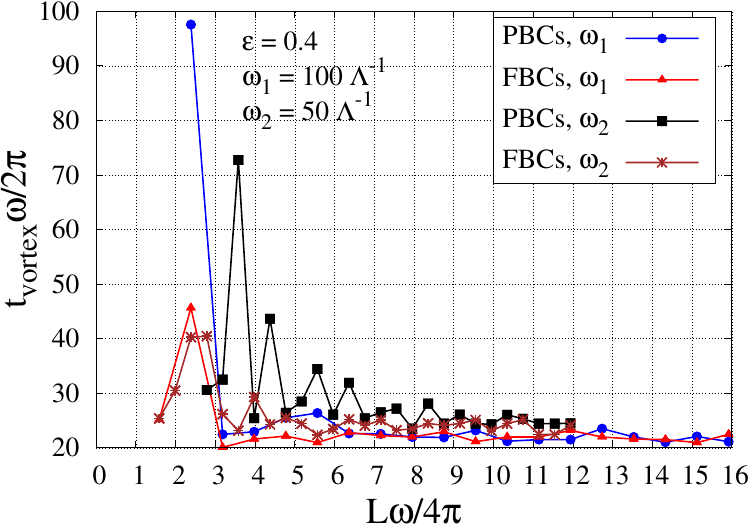}
 \caption{Figure shows the effects of using different boundary conditions on $t_{vortex}$ at 
 different system sizes $L$ and at two different $\omega$, $\omega_1$=100 $\Lambda^{-1}$ and 
 $\omega_2$=50 $\Lambda^{-1}$, where for each curve $\omega$ and $\varepsilon$ are fixed. The 
 parameters of effective potential are $\Phi_0$=10 $\Lambda^{-1}$ and $\lambda$=40. Blue(circle) 
 and black(square) lines correspond to simulations using periodic boundary conditions (PBCs) at 
 frequencies $\omega_1$ and $\omega_2$, respectively, while red(triangle) and brown(star) lines 
 correspond to simulations using fixed boundary conditions (FBCs), again, at frequencies $\omega_1$ 
 and $\omega_2$, respectively. In each curve, beyond the lowest value of $L\omega/4\pi$, $t_{vortex}$
 diverges.}
 \label{fig12}
\end{figure}

To investigate this in more detail, we have performed simulations at different system sizes $L$,
frequencies $\omega$, and boundary conditions. As mentioned earlier, in our simulations, 
we use periodic boundary conditions (PBCs). These boundary conditions affect our simulation 
results very strongly in the regime where finite size effects dominate. In general, we observe 
finite size effects when frequency of spacetime oscillations becomes $\omega$$\lesssim$$4\pi/ L$ 
or equivalently when $L\omega/4\pi$$\lesssim$1. To show these effects explicitly, and effects 
of boundary conditions in general, we study effects of changing lattice size $L$ and choice of 
boundary condition on $t_{vortex}$ by keeping $\omega$ and $\varepsilon$ fix. We show below that 
values of $t_{vortex}$ using fixed boundary conditions (FBCs) are relatively less affected with 
finite size effects in comparison with simulations using PBCs. In Fig.\ref{fig12}, we plot 
$t_{vortex}~\omega/2\pi$ versus $L\omega/4\pi$ (both are dimensionless numbers) for two different 
boundary conditions and at two frequencies $\omega$, $\omega_1$=100 $\Lambda^{-1}$ and $\omega_2$=50 
$\Lambda^{-1}$, where for each curve $\omega$ and $\varepsilon$(=0.4) are fixed. The parameters of 
effective potential are $\Phi_0$=10 $\Lambda^{-1}$ and $\lambda$=40. In this figure, blue(circle) 
and black(square) lines correspond to simulations using PBCs at frequencies $\omega_1$ and $\omega_2$, 
respectively, while red(triangle) and brown(star) lines correspond to simulations using FBCs, again, 
at frequencies $\omega_1$ and $\omega_2$, respectively. Under the spacetime oscillations, the lowest 
cut-off for $L\omega/4\pi$ to generate field excitations is 1, which must also be a cut-off for 
formation of vortices in the system. Therefore when $L\omega/4\pi$ approaches 1 from higher values, 
one would expect a growth in $t_{vortex}~\omega/2\pi$ which must diverge at $L\omega/4\pi$$\lesssim$1. 
This is what we see in the figure, where in each curve, beyond the lowest value of $L\omega/4\pi$, 
$t_{vortex}$ diverges. 

In the figure, for sufficiently large values of $L\omega/4\pi$ (say, greater than $\sim$10), 
$t_{vortex}~\omega/2\pi$ is almost invariant with increasing value of $L\omega/4\pi$, and takes 
a constant value $\simeq$23. This constant value seems to be a universal number, which is 
independent from choice of boundary condition and $\omega$. On the other hand, when 
$L\omega/4\pi$ approaches 1 (say, becomes less than $\sim$10), $t_{vortex}~\omega/2\pi$ 
starts increasing due to finite size effects and diverges when $L\omega/4\pi$ reaches very 
close to 1 (say, $\simeq$2 for PBCs, and $\simeq$1.5 for FBCs). Note that in the case of FBCs, 
the deviation in $t_{vortex}~\omega/2\pi$ from such a universal value is less in comparison 
with PBCs. Also, there are small lattice sizes, for which vortices are not formed using PBCs, 
whereas they are formed using FBCs. Thus, with FBCs, vortices could be formed easily in 
comparison with PBCs. It should be appreciated that for both boundary conditions, the number 
of cycles of spacetime oscillations required to form the first vortex-antivortex pair in the 
system is approximately the same at $L\omega/4\pi$$\gg$1 and is independent of frequency $\omega$.
Thus, this study 
clearly indicates that there is a universal number of cycle of spacetime oscillations to form 
first vortex-antivortex pair in the system if there are no finite size effects. All in all, 
this investigation suggests that even with a small frequency $\omega$, field may get excited 
(transverse excitation) under the spacetime oscillations if the simulation is performed on a 
sufficiently large lattice ($L$$\gg$$4\pi/\omega$) and the field is evolved for a significantly 
long time ($t$$\gtrsim$$t_{vortex}$$\simeq$$46\pi/\omega$$\simeq$$\frac{3}{2}\frac{\pi^4}{\omega}$).

\begin{figure}
 \includegraphics[width=1.0\linewidth]{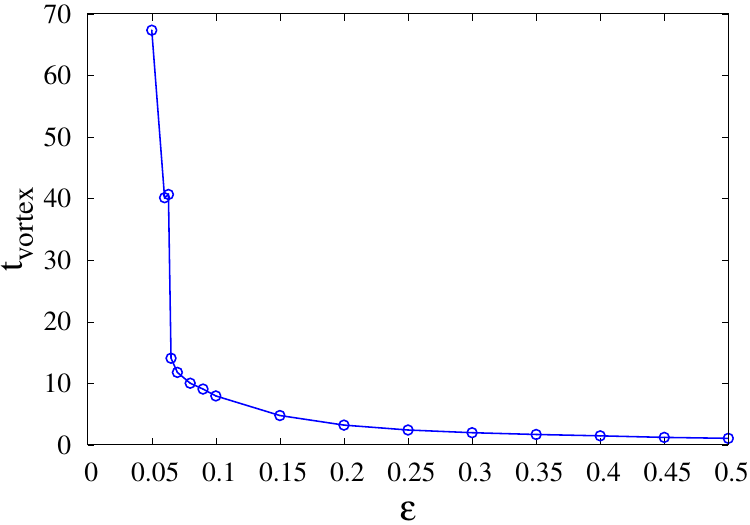}
 \caption{Figure shows the effect of $\varepsilon$ on $t_{vortex}$. Other parameters of simulations 
 are $\omega$=100 $\Lambda^{-1}$, $\Phi_0$=10 $\Lambda^{-1}$, and $\lambda$=40.}
 \label{fig13}
\end{figure}

In Fig.\ref{fig13}, we plot $t_{vortex}$ versus $\varepsilon$. Other parameters of 
simulations are $\omega$=100 $\Lambda^{-1}$, $\Phi_0$=10 $\Lambda^{-1}$, and $\lambda$=40. Figure 
clearly shows that decreasing $\varepsilon$ causes more time to form vortex-antivortex pair in the 
system. For lower values of $\varepsilon$, the system takes a longer time to achieve field excitations, 
where the large amplitude field oscillation stage persists for a longer time. In the regime of very 
small $\varepsilon$, the field excitations do not grow sufficiently to create vortices in the system. 
This behavior of field response to $\varepsilon$ is also a feature of {\it parametric resonance}, in 
which the resonance growth is strongly suppressed by decreasing the amplitude of time dependent 
parameter(s) of the oscillator \cite{landauMech}.  

\subsection{D. Formation of vortex-antivortex lattice under spacetime oscillations}

\begin{figure}
 \includegraphics[width=1.0\linewidth]{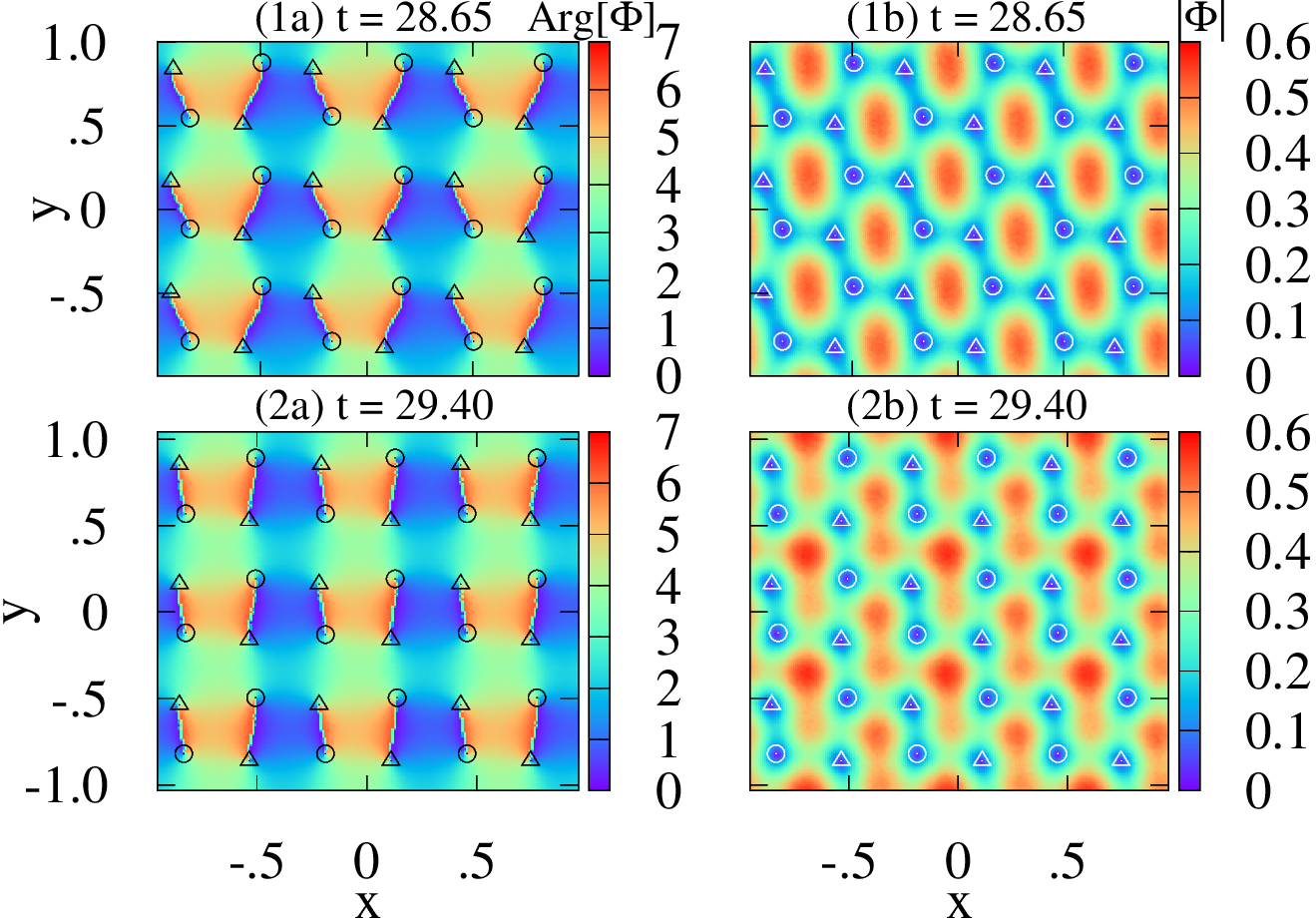}
 \caption{Figure shows the formation of vortex-antivortex lattice structures in the system. The locations of 
 vortex and antivortex in the plots are depicted by circles and triangles, respectively. The parameters of 
 the simulation are $\omega$=17 $\Lambda^{-1}$, $\varepsilon$=0.4, $\Phi_0$=0.1 $\Lambda^{-1}$,  and $\lambda$=4. 
 Plots (1a) and (1b) show the phase and magnitude of field $\Phi$ at time $t$=28.65 $\Lambda$, while plots (2a) 
 and (2b) show the same at time $t$=29.4 $\Lambda$.}    
 \label{fig14}
\end{figure}

The formation of vortex-antivortex lattice structure has interest in various condensed matter systems. 
In refs.\cite{sclatt1,sflatt1,sclatt2}, the formation of such lattices is studied in (i) superconducting 
films with magnetic pinning arrays, (ii) ultra-cold fermionic gases in two-dimensions, and (iii) superconducting 
twisted-bilayer graphene. Melting of the vortex-antivortex lattice in two-dimensional Fermi gases has been 
studied in ref.\cite{sflatt2}. For certain parameters of the simulation, we have also seen the formation 
of vortex-antivortex lattice structures under spacetime oscillations. In Fig.\ref{fig14}, we show 
the phase (left panel) and magnitude (right panel) of field $\Phi$ at two different times of field evolution. 
The parameters of the simulation are $\omega$=17 $\Lambda^{-1}$, $\varepsilon$=0.4, $\Phi_0$=0.1 $\Lambda^{-1}$, 
and $\lambda$=4. As mentioned earlier that, for $\omega$$\gg$$m_{\Phi}$, the formed vortices under spacetime
oscillations are not topological in nature. For the above parameters of the simulation also, $\omega$ is much greater 
than $m_{\Phi}$, therefore the formed vortices in Fig.\ref{fig14} do not satisfy the properties of topological 
vortices. The core size of these vortices is much less than $m_{\Phi}^{-1}$, rather it is given by 
$\sim$$\omega^{-1}$, where the field values outside these vortices are six times greater than VEV. 
Fig.\ref{fig14}(1a) and Fig.\ref{fig14}(1b) show the phase and magnitude of the field $\Phi$ at time $t$=28.65 
$\Lambda$, while Fig.\ref{fig14}(2a) and Fig.\ref{fig14}(2b) show the same at time $t$=29.4 $\Lambda$. On the 
plots, circles and triangles indicate the locations of vortices and antivortices, respectively. In the left 
panel, these locations correspond to +$2\pi$ and $-$$2\pi$ variations of phase of $\Phi$, respectively, and in 
the right panel, they correspond to the places where the magnitude of $\Phi$ is zero. One can clearly see 
that there are vortex-antivortex lattice structures in both, upper and lower, panels. We observe that these lattice 
structures keep on changing with time. Formation of these structures does not depend upon the amplitude of initial 
fluctuations, i.e. on the value of $\beta$. It may be possible that such lattice structures are arising because of 
the use of periodic boundary conditions in the simulation.  

\section{V. Conclusion and future directions}

In this work, we have performed (2+1)-dimensional simulations for a complex scalar field in the presence of oscillating 
spacetime metric background. We have considered spacetime metric as a periodic time-dependent perturbations on the top of 
Minkowski metric. The field is taken on the VEV of U(1) symmetry broken effective potential, where small random fluctuations 
in the initial field configuration have been considered. We find that depending upon the amplitude and frequency of metric 
oscillations, field undergoes {\it parametric resonance}, which leads to the generation of field excitations and the 
formation of vortex-antivortex pairs in the system. There is a fundamental difference in the evolution of field in the 
frequency regimes $\omega$$>$$m_\Phi$ and $\omega$$\leq$$m_\Phi$. In the low frequency regime mainly transverse excitation 
arises, and well separated vortex-antivortex pairs form in the system, whereas in the high frequency regime, longitudinal 
excitation of the field also arises prominently.  

In our simulation, finite size effects restrict the generation of field excitations in the regime $L\omega/4\pi$$\lesssim$1, 
where $L$ is the system size, and $\omega$ is the frequency of spacetime oscillations. Our study suggests that, in principle, 
any low frequency can excite the field if a suitable system size is chosen ($L$$\gg$$4\pi/\omega$) and the field is evolved for 
a significantly long time ($t$$\gtrsim$$\frac{3}{2}\frac{\pi^4}{\omega}$). We have studied the effects of frequency, boundary 
conditions, and system size on the {\it number of cycle} of spacetime oscillations required to form first vortex-antivortex 
pair in the system. This suggests that there is a universal number of cycle of spacetime oscillations for formation of first 
vortex-antivortex pair in the system if there are no finite size effects. The formation of vortices also depends upon the 
amplitude of spacetime oscillations; the vortex formation time increases by decreasing amplitude. For certain parameters of 
the simulation, we have seen the formation of vortex-antivortex lattice structures. However, such vortices do not satisfy the 
properties of topological vortices as in such case, $\omega$ is much greater than $m_{\Phi}$ (mass of the field), for which 
the generated field-modes have much larger momentum in comparison with $m_{\Phi}$. Only in the case of 
$\omega$$\lesssim$$m_{\Phi}$, the formed vortices under spacetime oscillations are topological in nature.

In this work, we have ignored an important aspect of the system evolution, which is the back-reaction of energy density variations 
of field on the spacetime metric. As we have shown, under spacetime oscillations, vortices are formed, which have an energy density 
profile, with a maximum value in the vortex core. Therefore, depending upon the energy of vortex 
configuration, the spacetime manifold itself may be affected by these energy density variations, which may affect the further 
evolution of the field. It would be interesting to see the field evolution under such a complete scenario. We will try to pursue this 
in future. 

To study the time evolution of condensate field of neutron stars superfluidity during BNS merger, a full (3+1)-dimensional simulation 
with an appropriate time dependent deformation of star metric is required. It will reveal whether the time scale and length scale of 
the whole process are sufficient to excite the condensate field to lead the formation of vortex-antivortex pairs in the interior 
superfluid. Our present study suggests that there is no possibility of generation of condensate excitations of neutron star superfluidity 
under the phenomenon of parametric resonance due to finite size effects. However, a detailed investigation is needed to reveal whether 
there is any other method of generation of such excitations.  

The present work has applications in other systems also. The {\it second homotopy group} of $S^2$-manifold is non-trivial, i.e. 
$\pi_2(S^2)$=$\mathcal{Z}$ \cite{mermin}, which allows the existence of {\it topological texture} in (2+1)-dimensions (baby 
Skyrmion) and {\it topological monopole} in (3+1)-dimensions. Similarly, $S^3$-manifold (group space of $SU(2)$) allows the 
existence of {\it topological texture} in (3+1)-dimensions as the {\it third homotopy group} is non-trivial, i.e. 
$\pi_3(S^3)$=$\mathcal{Z}$ \cite{mermin}. These topological objects have cosmological interests, the formation of which could 
be studied under spacetime oscillations as well (similar to the formation of vortices studied in this paper). In the 
{\it abelian-Higgs model} and pure $U(1)$ gauge theory also, one could study the formation of flux-tubes and generation of 
magnetic field, respectively, under these oscillations. 

This study can also be extended to U(1) symmetry broken effective potential with a small explicit symmetry breaking term. In such a
case, there must be a lowest frequency cut-off of spacetime oscillations for the generation of field excitations set by the mass of 
pseudo-Goldstone modes; only frequencies higher than or equal to the mass of these modes can lead to the field excitations. In our
simulation, we have seen the generation of field excitations in the presence of such a term (we have not presented this here). This 
result suggests a possibility of generation of excitations in the {\it axion-like} field having mass $\sim$$10^{-22}$ - $10^{-21}$ 
eV \cite{axion} and coupled with spacetime oscillations. The {\it axion-like} particles having above mass range are considered to be 
a possible candidate of dark matter, known as the $\textquoteleft$wave dark matter' or $\textquoteleft$fuzzy dark matter' \cite{axion}. 
To excite this field, correspondingly, the frequency of spacetime oscillations must be $\gtrsim$ 0.1 - 1 {\textmu}Hz. This frequency 
can be easily achieved by BNS \cite{bns1} and black-hole merger \cite{blackhole} systems, which produce gravitational waves (GW) in 
the frequency range of $\sim$10Hz - 1kHz with a significantly large strain amplitude. We expect that a sustained spacetime oscillations 
will give rise excitation in {\it axion-like} field over a length scale having momentum of field corresponding to the frequency of 
spacetime oscillations. The effect will be more pronounced in the nearest region of GW source as the strain amplitude is significantly 
large there. If such excitation arises, then as soon as GWs pass completely, the field will start rolling toward the minima of 
effective potential and perform coherent oscillations at various constant-$r$ hypersurfaces about the source. We will present this 
phenomenon in detail in our future work.

\section{acknowledgments}
 We are very grateful to Ajit M. Srivastava and Subhroneel Chakrabarti for very useful discussions, suggestions, and comments on the manuscript. We 
 also would like to thank A.P. Balachandran, P.S. Saumia, Abhishek Atreya, T.P. Sreeraj, Nirupam Dutta, and Minati Biswal for useful discussions and 
 comments on the work.


\begin{thebibliography}{99}

\bibitem{zrk1} W.H. Zurek, Phys. Rep. {\bf 276}, 177 (1996).  

\bibitem{rjnt} A. Rajantie, Int. J. Mod. Phys. A {\bf 17}, 1 (2002). 

\bibitem{mermin} N.D. Mermin, Rev. Mod. Phys. {\bf 51}, 591 (1979). 

\bibitem{kbl} T.W.B. Kibble, J. Phys. A {\bf 9}, 1387 (1976); Phys. Rep. {\bf 67}, 183 (1980).

\bibitem{skyrm} V.S. Kumar, B. Layek, A.M. Srivastava, S. Sanyal, and V.K. Tiwari, Int. J. Mod. Phys. A {\bf 21},
1199 (2006).

\bibitem{bias} S.S. Dave and A.M. Srivastava, Europhys. Lett. {\bf 126}, 31001 (2019).

\bibitem{landau} L.D. Landau and E.F. Lifshitz, {\bf Statistical Physics Part 2}, Pergamon Press Ltd. (1980).

\bibitem{tilley} D.R. Tilley, J. Tilley, {\bf Superfluidity and Superconductivity}, 3rd ed. (Graduate Student Series 
in Physics, Institute of Physics Publishing, Bristol and Philadelphia, 1990).

\bibitem{resodef} S. Digal, R. Ray, S. Sengupta, and A.M. Srivastava, Phys. Rev. Lett. {\bf 84}, 826 (2000).

\bibitem{kink} R. Ray and S. Sengupta, Phys. Rev. D {\bf 65}, 063521 (2002).

\bibitem{oscBEC1} E.A.L. Henn, J.A. Seman, G. Roati, K.M.F. Magalh$\tilde{a}$es, and V.S. Bagnato, Phys. Rev. Lett. {\bf 103}, 045301 (2009);
E.A.L. Henn et al., Phys. Rev. A {\bf 79}, 043618 (2009).

\bibitem{oscBEC2} V.I. Yukalov, A.N. Novikov, and V.S. Bagnato, Phys. Lett. A {\bf 379}, 1366 (2015).

\bibitem{resoBEC1} V.I. Yukalov, E.P. Yukalova, and V.S. Bagnato, Phys. Rev. A {\bf 66}, 043602 (2002).

\bibitem{bns1} B.P. Abbott et al. (LIGO Scientific Collaboration and Virgo Collaboration), Phys. Rev. Lett. {\bf 119}, 161101 (2017).

\bibitem{ns1} R.E. Packard, Phys. Rev. Lett. {\bf 28}, 1080 (1972).

\bibitem{ns2} D.G. Yakovlev, A.D. Kaminker, and K.P. Levenfish, Astron. Astrophys. {\bf 343}, 650 (1999).

\bibitem{krishna} M.G. Alford, A. Schmitt, K. Rajagopal, and T. Sch$\ddot{a}$fer, Rev. Mod. Phys. {\bf 80}, 1455 (2008).

\bibitem{proposal} {\bf Gribov-80 Memorial Volume; Quantum Chromodynamics And Beyond}, C.N. Colacino, page 449-454, World Scientific (2010).

\bibitem{recentMerger} A. Perego, S. Bernuzzi, and D. Radice, Eur. Phys. J. A {\bf 55}, 124 (2019).

\bibitem{nsSFtc} L. Amundsen and E. {\O}stgaard, Nucl. Phys. A {\bf 437}, 487 (1985); Nucl. Phys. A {\bf 442}, 163 (1985).

\bibitem{anandan} J. Anandan, Phys. Rev. Lett. {\bf 47}, 463 (1981).

\bibitem{introSF} A. Schmitt, {\bf Introduction to Superfluidity; Field-theoretical Approach and Applications}, Springer (2015).

\bibitem{relSF1} C. Xiong, M.R.R. Good, Y. Guo, X. Liu, and K. Huang, Phys. Rev. D {\bf 90}, 125019 (2014).

\bibitem{carroll} S. Carroll, {\bf Spacetime and Geometry; An Introduction to General Relativity}, Addison Wesley (2004).

\bibitem{landauMech} L.D. Landau and E.F. Lifshitz, {\bf Mechanics; Third Edition}, Elsevier (2007).

\bibitem{geodesic} M.J. Bowick, L. Chandar, E.A. Schiff, and A.M. Srivastava, Science {\bf 263}, 943 (1994); [hep-ph/9208233].

\bibitem{sclatt1} M.V. Milo$\check{s}$evi$\acute{c}$ and F.M. Peeters, J. Low Temp. Phys. {\bf 139}, 257 (2005). 

\bibitem{sflatt1} S.S. Botelho and C.A.R. S$\acute{a}$ de Melo, Phys. Rev. Lett. {\bf 96}, 040404 (2006).

\bibitem{sclatt2} Y. Su and S.-Z. Lin, Phys. Rev. B {\bf 98}, 195101 (2018).

\bibitem{sflatt2} G. Cao, L. He, and X.-G. Huang, Phys. Rev. A {\bf 96}, 063618 (2017).

\bibitem{axion} L. Hui, J.P. Ostriker, S. Tremaine, and E. Witten, Phys. Rev. D {\bf 95}, 043541 (2017).

\bibitem{blackhole} B.P. Abbott et al. (LIGO Scientific Collaboration and Virgo Collaboration), Phys. Rev. Lett. {\bf 116}, 061102 (2016).

\end{thebibliography}
\end{document}